\documentclass[referee]{aa} 
\usepackage{graphicx}
\usepackage{natbib}
\bibpunct{(}{)}{;}{a}{}{,} 
\newcommand{\lsim}{\ \raise -2.truept\hbox{\rlap{\hbox{$\sim$}}\raise 5.truept\hbox{$<$}\ }}
\newcommand{\gsim}{\ \raise -2.truept\hbox{\rlap{\hbox{$\sim$}}\raise 5.truept\hbox{$>$}\ }}
\begin{document}
\title{Pulsation properties of C stars in the Small Magellanic Cloud}

   \author{Gabriella Raimondo\inst{1,2}
          \and
          Maria-Rosa L.\ Cioni\inst{1}
      \and
          Marina Rejkuba\inst{1}
      \and
          David R.\ Silva\inst{1}
          }

   \offprints{G. Raimondo, \email{raimondo@te.astro.it}}

   \institute{European Southern Observatory, Karl-Schwarzschild-Str.\ 2, D-85748
    Garching bei M\"{u}nchen, Germany \\
       E-mail: graimond@eso.org, mcioni@eso.org, mrejkuba@eso.org, dsilva@eso.org
    \and
    INAF, Osservatorio Astronomico di Teramo, Via M. Maggini, I-64100, Teramo, Italy}
   \date{Received date / Accepted date}

    \authorrunning{Raimondo et al.}
    \titlerunning{C-stars in the SMC}

\abstract{ A sample of carbon-rich stars (C-stars) in the Small
Magellanic Cloud (SMC) was selected from the combined 2MASS and
DENIS catalogues on the basis of their $J-K_s$ colour.  This
sample was extended to include confirmed C--stars from the
Rebeirot et al.\ (1993) spectroscopic atlas. In this combined
sample (N = 1152), a smaller number (N = 1079) were found to have
MACHO observations. For this sub--sample, light curves were
determined and 919 stars were found to have high quality
light-curves with amplitudes of at least 0.05 mag. Of these stars,
only 4\% have a well--defined single period -- most of these have
multiple well-defined periods, while 15\% have highly irregular
light--curves. The distribution of the logarithm of the period
versus magnitude, colour, period ratio (if applicable), and
amplitude was analyzed and compared with previous works. Variable
C-stars are distributed in three sequences: B, C, and D from Wood
et al. (1999), and do not populate sequences with periods shorter
than $\log P \sim 1.5$.  Stellar ages and masses were estimated
using stellar evolutionary models.
    }
  \maketitle
%

\section{Introduction}

Asymptotic Giant Branch (AGB) stars can be separated into two
classes based on their spectra: oxygen--rich (O--rich or M--stars)
and carbon--rich stars (C--rich or C--stars). M--stars have more
oxygen than carbon in their atmospheres ($C/O<1$), while C--stars
display $^{12}C$ enrichment ($C/O>1$) due to dredge--up caused by
thermal pulsation \citep{Iben&Renzini83}.  These thermal pulses
also lead to mass loss, as well as to luminosity variations with
periods of $\sim$100 days or longer and peak-to-peak amplitude
variations up to a few magnitudes at visual wavelengths.

All stars are oxygen--rich when they enter the AGB phase. Whether
or not they become C--stars depends primarily on the efficiency of
the third dredge--up and the extent and time-variation of the
mass-loss \citep[e.g.\ ][]{Iben81, Marigo+99}. In metal--poor
stars, fewer $^{12}C$ atoms are necessary to change the envelope
from oxygen to carbon dominated ($C/O>1$); therefore, fewer
thermal pulses are needed to convert an M--star into a C--star.
Conversely, mass loss is expected to be stronger in metal-rich
stars, leading to shorter AGB and C--star phases.

In the past, it was thought that both oxygen--rich and
carbon--rich Mira variables follow a well--determined
period--luminosity (PL) relation in the near--IR regardless of the
host system mean metallicity or type, e.g.\ in a globular cluster
\citep{Feast+02}, dwarf galaxy \citep{Glass&Lloydevans81}, spiral
galaxy \citep{Glass+95,Vanleeuwen+97}, or elliptical galaxy
\citep{Rejkuba04}. However, the availability of long--term
photometric monitoring data provided by the microlensing observing
projects, e.g.\ MACHO \citep{Alcock+92}, OGLE \citep{Zebrun+01,
Udalski+97}, and EROS \citep{Aubourg+93}, and large-scale
near-infrared (NIR) photometric surveys, like the Two Micron All
Sky Survey \citep[2MASS, ][]{Skrutskie+97} and the Deep
Near--Infrared Southern Sky Survey \citep[DENIS, ][]{Epchtein+97}
have opened a new window on this issue and revealed that AGB stars
lie on multiple parallel sequences in the PL diagram (Cook et al.
1997; Wood et al. 1999). Furthermore, red giant branch (RGB) stars
at the tip of the RGB were also found to vary (Ita et al. 2002,
Kiss \& Bedding 2003). These results have provided new and
significant constraints for theoretical pulsation models.

Differences between O--rich and C-rich Long--Period Variables
(LPVs) have also been found.  Using a small sample of AGB stars in
the SMC observed by the Infrared Space Observatory (ISO), 2MASS,
DENIS and MACHO, Cioni et al.\ (2003) concluded that the period
distribution of C--stars peaks at about 280 days. They also noted
that C-stars have a larger amplitude with respect to M-stars,
contrary to what was derived for the LMC AGB stars, where both
types showed a similar amplitude distribution \citep{Cioni+01}.
Studying a much larger sample of C and M LPVs in both Magellanic
Clouds, Ita et al.\ (2004b) confirmed that O-- and C--rich Miras
follow different period $vs.$ ($J-K$) colour relations
\citep{Feast+89}, that C--rich Miras tend to have greater
$I$--band amplitudes at redder $J-K$ colour, and that the
amplitudes of O--rich Miras are independent of colour. Groenewegen
(2004, G04) reached similar conclusions.

However, the studies by Cioni et al. (2001) and Ita et al. (2002)
were limited to LPVs with $P < 1000$ days, since the OGLE-II
observations only span a time--baseline of about 1200 days.  More
recently, Fraser et al.\ (2005) presented an analysis of the eight
year light-curve MACHO data for LPVs in the LMC and found that
C-stars occupy only two of the sequences in the period-luminosity
diagram. Furthermore, dust-enshrouded stars are located in the
high-luminosity ends of the both sequences.

In this paper, we have extended the work of Cioni et al.\ (2003)
and complemented the work of Fraser et al. (2005) by investigating
the variability properties of all C--stars observed by MACHO in
the SMC.  We used the Master Catalogue of stars toward the
Magellanic Clouds ($\mathrm{MC}^2$) by Delmotte et al.  (2002) and
Delmotte (2003) to identify C-stars in the SMC.  This catalogue
provides a cross-correlation between the DENIS Catalogue towards
the Magellanic Clouds (DCMC -- $IJK_s$) and the 2nd Incremental
Release of the 2MASS point source catalogue ($JHK_s$) covering the
same region of the sky. C--stars were selected statistically on
the basis of their red $J-K_s$ colours ($J-K_s \geq 1.33$; Cioni
et al. 2003). This selection was checked through cross-correlation
with the Rebeirot et al.\ (1993, hereafter RAW93) catalogue of
spectroscopically confirmed C-stars
(Sect.~\ref{subsection:photselection}).  C--stars found
spectroscopically by RAW93 but with $J-K_s < 1.33$ were later
included in the sample.  In Sect.~\ref{section:photdata}, we
present the selection of C-stars and the extraction of the
corresponding light--curves from the MACHO database.  Section
\ref{section:analysis} discusses the analysis and resulting
light-curve parameters.  The ($logP$, $K_s$), ($J-K_s$, $K_s$) and
other diagrams are discussed in Section~\ref{section:discussion},
while Section~\ref{section:conclusion} concludes this work.
Details about the method developed to determine periods and
amplitudes and the quality assessment of the data and of the
relevant parameters are given in Appendix A.

\section{Photometric Data and Light--Curves}
\label{section:photdata}
\subsection{C--stars photometric selection}
\label{subsection:photselection}

\begin{figure}
\center
  \includegraphics[width=9cm]{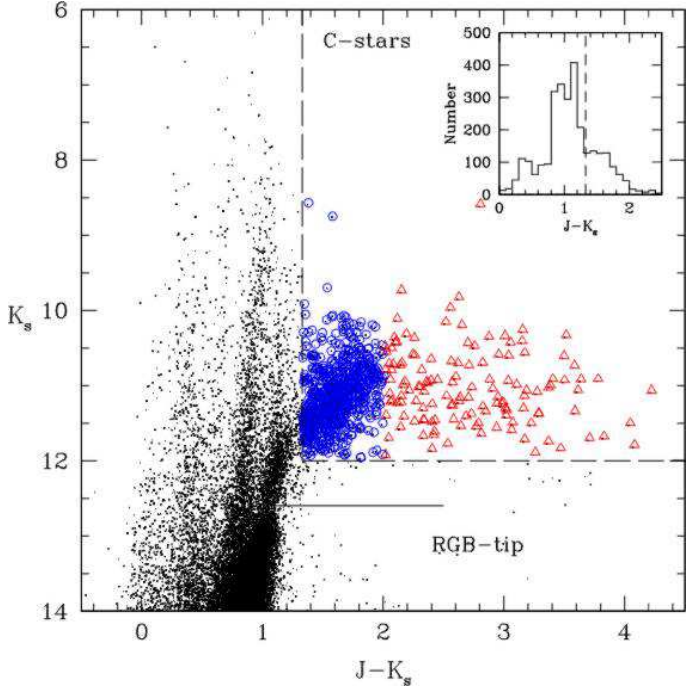}
   \caption{Near--IR CMD of SMC stars from 2MASS within the observed
   MACHO fields (dots). The C--stars region is marked by two perpendicular
   dashed lines: C--stars with $J-K_s \leq 2$ mag are indicated with blue
   circles; red triangles refer to obscured AGB stars
   ($J-K_s>2$ mag). The position of the RGB--tip is also indicated.
   In the upper right corner we show the number of sources with $K_s\leq12$ vs.
   $J-K_s$ colour; note that the branch of C--stars is well separated. See
   the electronic edition of the Journal for a colour version of the
   figure.}  \label{fig:cmd}
\end{figure}

\begin{figure}
  \includegraphics[width=9cm]{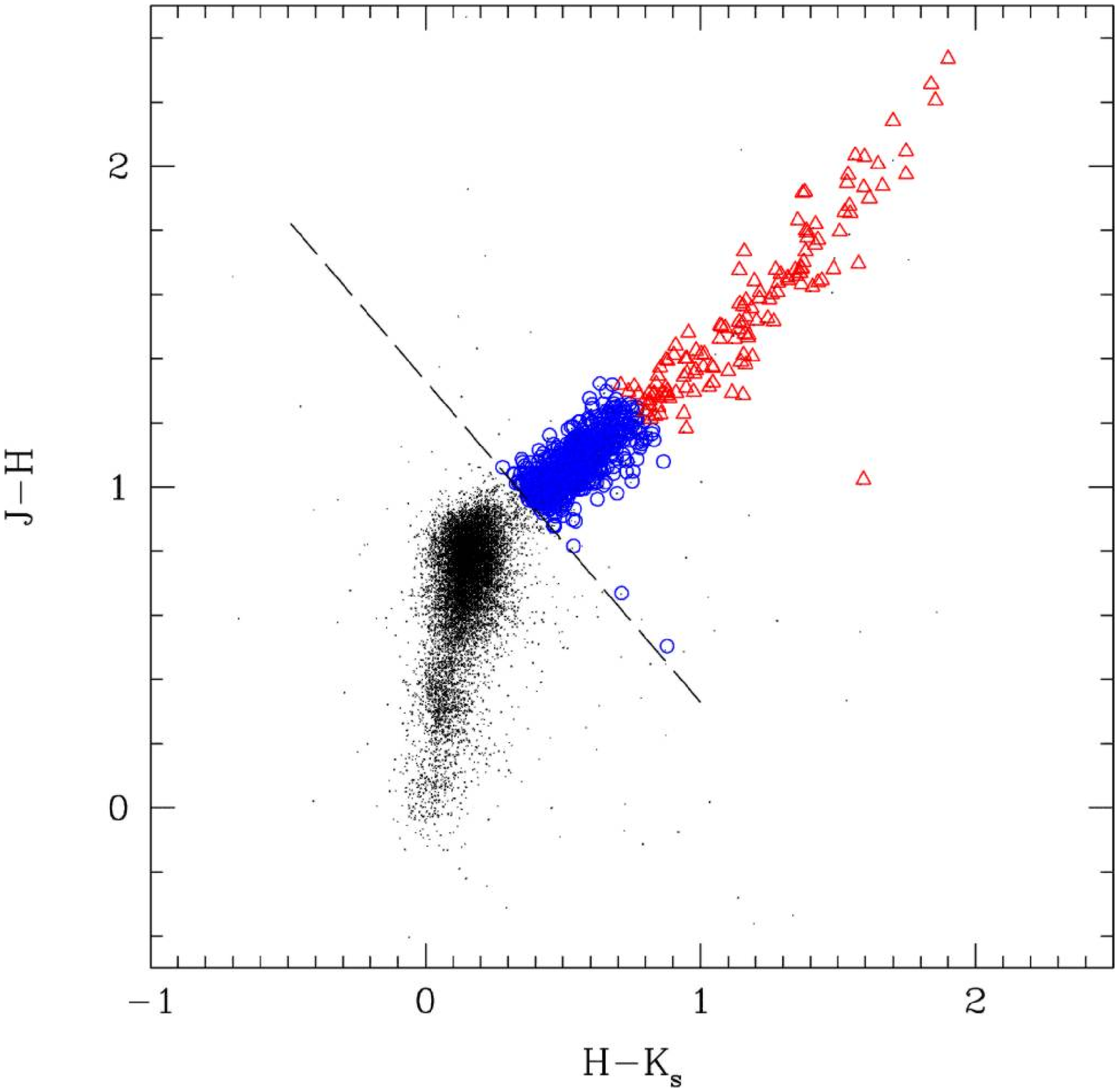}
   \caption{$J-H$ $vs.$ $H-K_s$ diagram of SMC stars
   within the MACHO fields with $K_s < 14$ mag. The dashed line corresponds
   to $J-K_s=1.33$ mag. Symbols are as in Fig.~\ref{fig:cmd}.
   See the electronic edition of the Journal for a colour version of the
   figure. }
   \label{fig:jhhk+ikjkdcmc}
\end{figure}

The $\mathrm{MC}^2$ catalogue, containing the cross--correlation
between DENIS and 2MASS surveys, as well as optical UCAC1 and
GSC2.2 catalogues toward the LMC, was published by Delmotte et al.
(2002). Here, we use its extension to the SMC (Delmotte 2003) and
the near-IR information of the catalogue only. The region
confidently populated by C--stars ($J-K_s \geq 1.33$ mag and $K_s
\leq 12$ mag) contains a total of 1657 stars. 805 stars within the
MACHO fields satisfy the photometric criterion. The MACHO project
observed 6 fields (each of 0.49 deg$^2$), covering the densely
populated bar of the SMC of approximately 3 square degrees in
total.

Figure \ref{fig:cmd} shows the near--IR $K_s$ $vs.$ $J-K_s$
colour--magnitude diagram (CMD) of stars from $\mathrm{MC}^2$ and
within the MACHO fields. Evolved AGB stars occupy the region above
the RGB--tip at $K_s \la 12.6$ mag \citep{Cioni+00} and $J-K_s
\gsim 0.97$ mag.  At $K_s\sim 12$ mag the split into two branches
is significant, though it starts already at $K_s\sim 12.5$.
C--stars populate the well--extended tail toward red colours,
while M--stars lie along the almost vertical sequence at $ J-K_s
\sim 1.2$ mag. They reach a maximum luminosity of $K_s\simeq 10$
mag, except for 3 stars that have $K_s\simeq 9$ and are likely to
be more massive O-rich stars (G04). Thus, these 3 have been
excluded from the sample. Dust--enshrouded AGB stars are located
at redder colours ($J-K_s \gsim 2$ mag). These obscured stars can
be either C--rich or O--rich, and spectra are needed to
distinguish between the two types. We include them in our
analysis, unless they are explicitly rejected by RAW93 (see
discussion of contamination by O--rich stars below). The small box
in the upper right corner shows the histogram of sources with
$K_s<12$ $ vs.$ $J-K_s$ colour. Indeed, the branch of C--stars is
well separated from O--rich stars at about $J-K_s=1.33$. Other
approximately vertical sequences in the main figure at $J-K_s\leq
0.97$ are populated by either foreground galactic stars or red
supergiants and upper main--sequence stars that belong to the SMC
(i.e. Nikolaev \& Weinberg 2000).

Figure \ref{fig:jhhk+ikjkdcmc} displays the $J-H$ $vs.$ $H-K_s$
colour--colour diagram of the SMC sources within the MACHO fields.
This diagram is useful for identifying stars with large infrared
excess \citep{Bessell&Brett88, Nikolaev&Weinberg00}. The main
feature in the diagram is an extended branch to the right side of
the dashed line that marks our selection of C--stars ($J-K_s=1.33$
mag).  It has been recognized as the locus of TP--AGB stars (e.g.
Marigo et al. 2003). Circles and triangles, as in
Fig.~\ref{fig:cmd}, correspond to stars in our sample. The less
populated region at $J-H \gsim 1.2$ and $H-K_s \gsim 0.8$ mag
corresponds to obscured AGB stars.  Along this branch a small
contamination of stars with $K_s>12$ mag is present (small dots).
A handful of them might also be C--stars with very large
amplitudes caught at minimum light or faint extrinsic C--stars
(Westerlund et al. 1995).

\subsection{C--stars confirmed spectroscopically}

\begin{figure*}
\center
   \includegraphics[width=5.5cm]{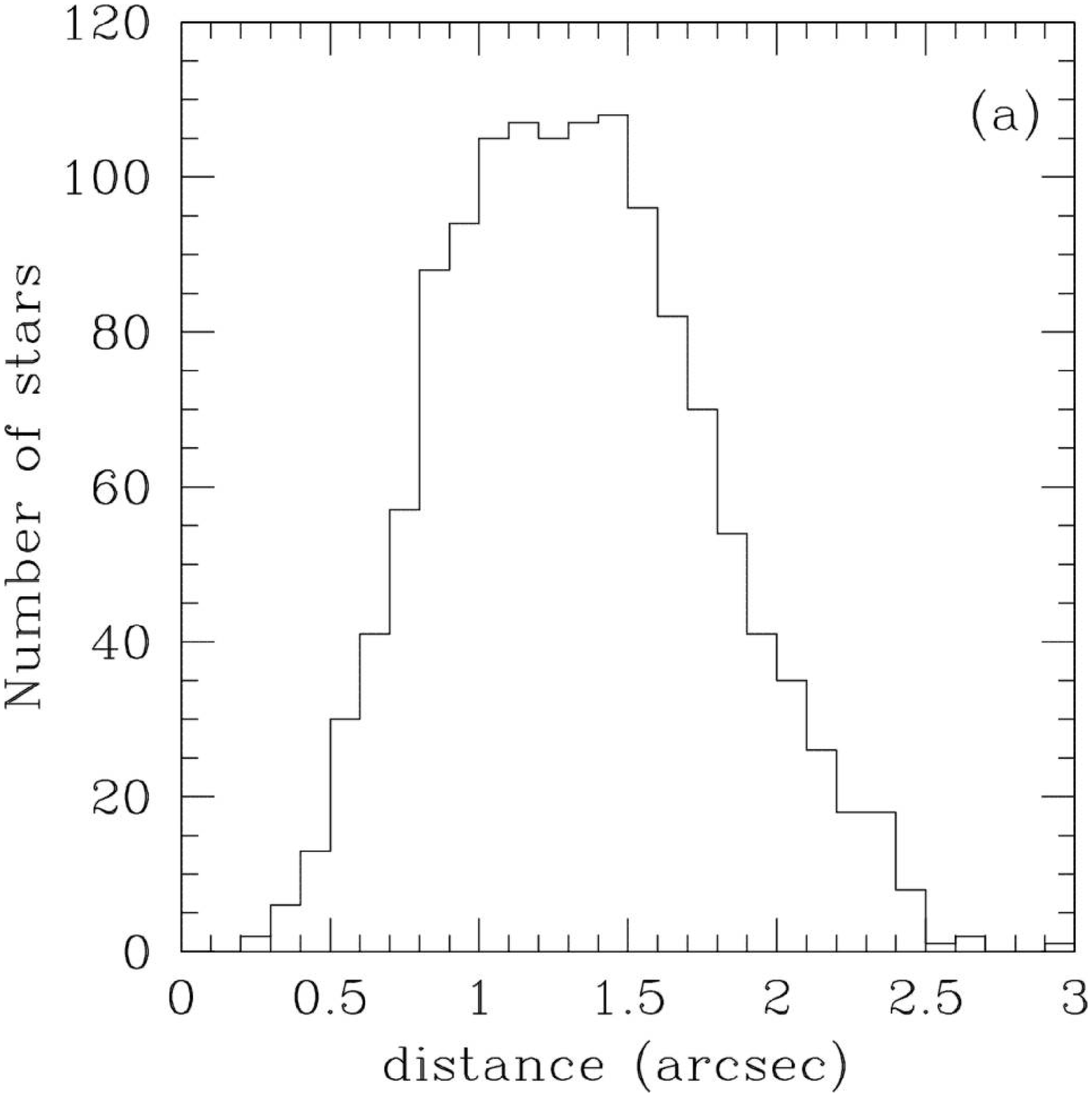}
   \includegraphics[width=5.5cm]{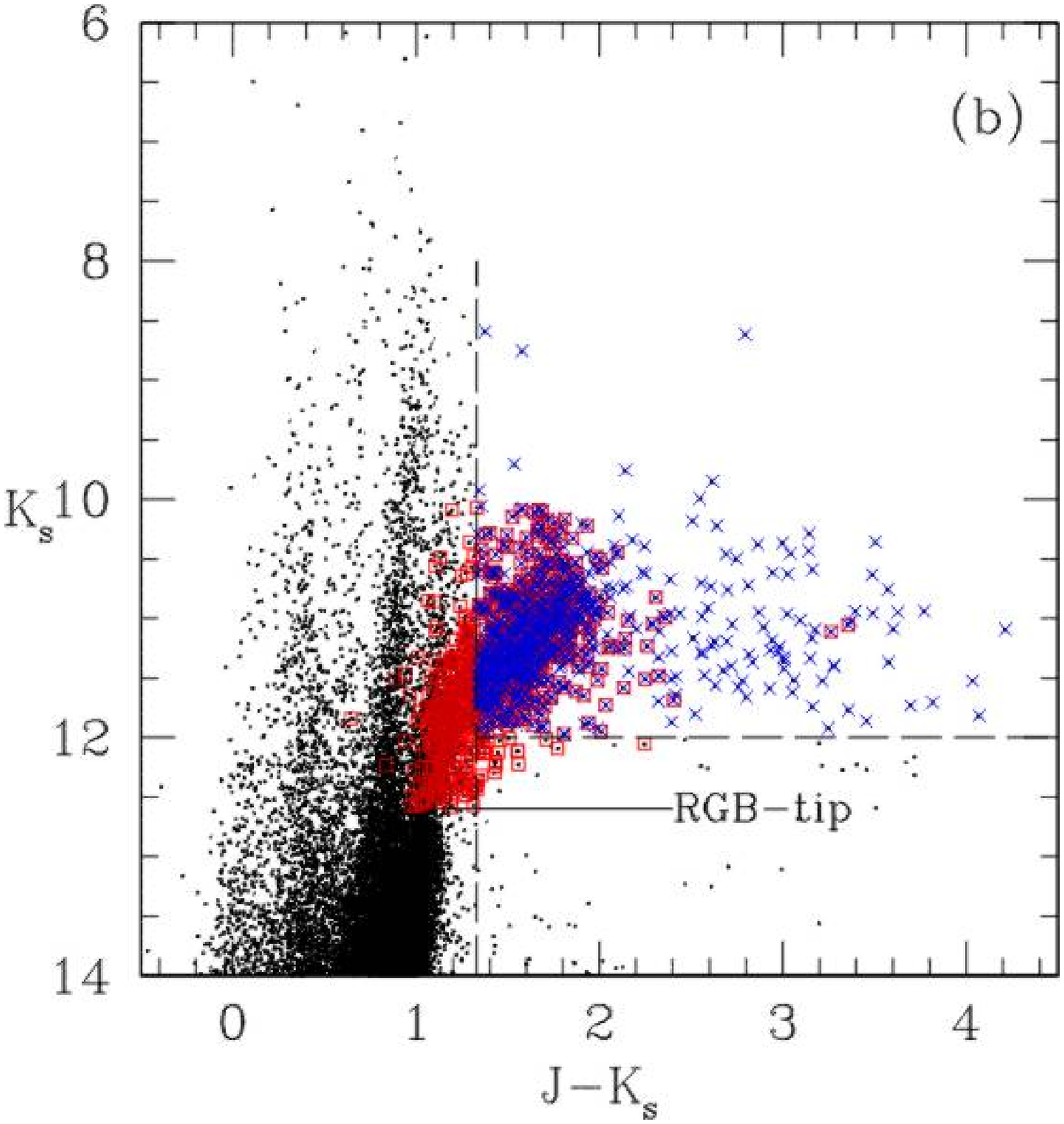}
   \includegraphics[width=5.5cm]{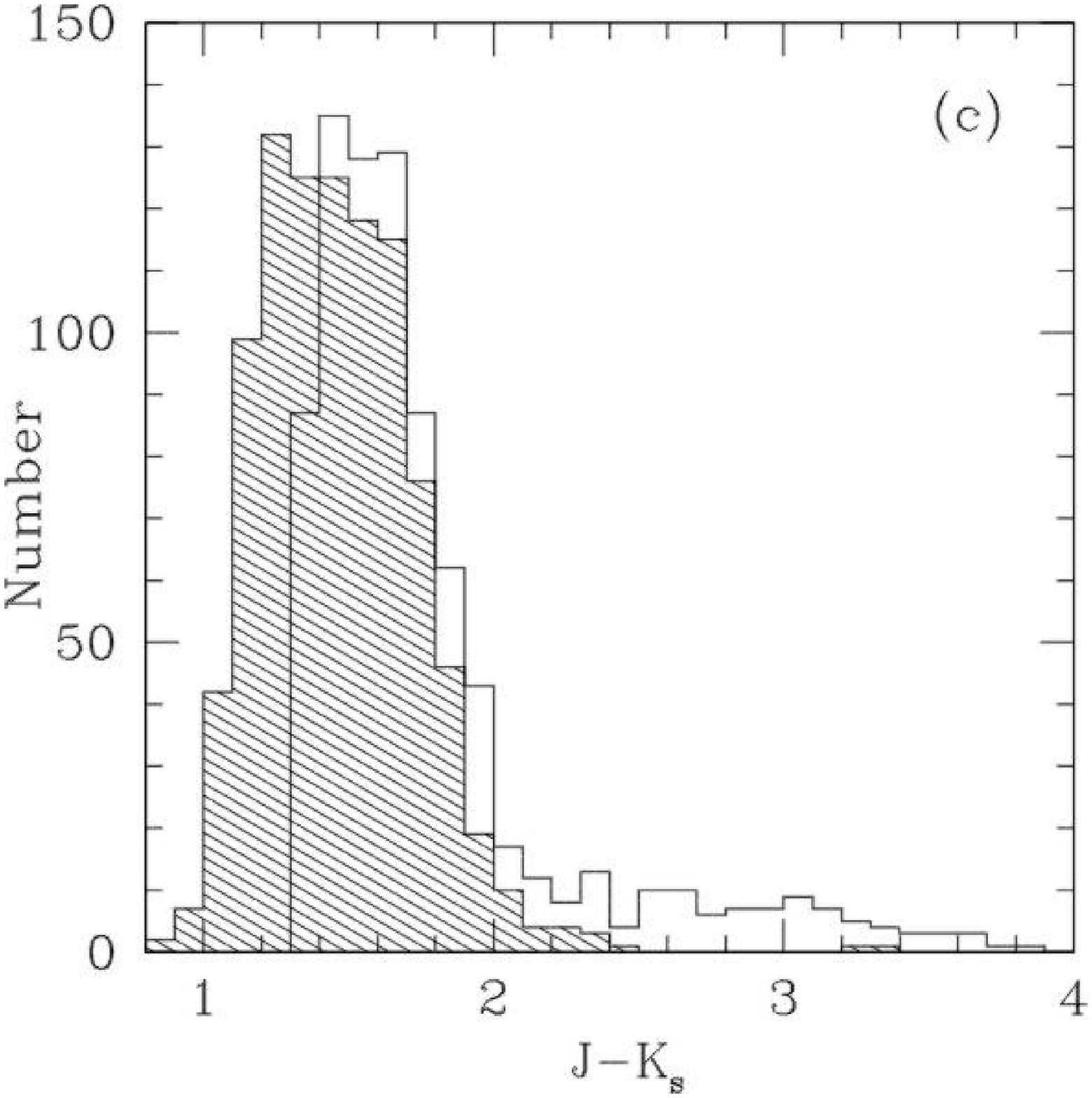}
   \caption{{\bf Panel $(a)$}: Absolute distance between 2MASS
   and RAW93 stars in the $\mathrm{MC}^2$ catalogue. {\bf
   Panel $(b)$}: $K_s$ $vs.$ $J-K_s$ CMD of all SMC stars in the MACHO
   field (small black dots). C-stars included in the RAW93 catalogue and
   within the  MACHO fields are plotted as red squares. Blue crosses represent
   the sample of photometrically selected C-stars. Near--IR photometry comes
   from the 2MASS survey. {\bf Panel $(c)$}: histogram of C--stars
   in our photometrically selected
   sample (empty) and in the RAW93 catalogue (dashed).
   See the electronic edition of the Journal
   for a colour version of the figure.}  \label{fig:rebeirot}
\end{figure*}

Our photometric selection criteria of C--stars were checked against
the spectroscopically confirmed C--stars in the RAW93 catalogue.
Among 1707 stars listed by RAW93 we rejected 27 that were classified
by the authors as ``doubtful'' (flag=1 in Column 8 of their
Table~4). Then, we cross--correlated the coordinates of the remaining
1680 C--stars with stars in the entire $\mathrm{MC}^2$ catalogue
brighter than the RGB tip (i.e.  $K_s=12.6$ mag), by adopting a
searching radius of 3\arcsec.  The absolute distance between the 2MASS
counterpart of the RAW93 sources is shown in
Fig.~\ref{fig:rebeirot}$a$. We found that 1275 C--stars in the RAW93
catalogue have a 2MASS counterpart within 3\arcsec, and the peak of
the distance distribution is at $d \sim 1.2$\arcsec.  The majority of
the spectroscopically confirmed C--stars are well matched to stars in
our sample when the same colour and $K_s$--magnitude criteria are
adopted \citep[see also][]{Cioni+00}. Restricting the area to those
fields observed by MACHO, there are 931 C--stars in common between
RAW93 and $\mathrm{MC}^2$.  The location of these stars in the
near--IR CMD is presented in Fig.~\ref{fig:rebeirot}$b$.

We find that 73\% of 802 C--stars photometrically selected have
confirmed C--type spectra. Our photometric selection identifies
more C--stars at $J-K_s \gsim 2$ mag with respect to RAW93. This
might be because 1) some stars with these colours might be O--rich
AGB stars, and 2) some obscured stars were probably below the
detection limit in the RAW93 survey. Due to metallicity dependence
of the C--star life--time, in a population with the metallicity of
the SMC we expect only few of our non-RAW93 stars to belong to the
first category. Many spectroscopically confirmed C--stars have a
colour bluer than $J-K_s=1.33$ mag and a magnitude fainter than
$K_s=12$ mag, overlapping the region where O-rich stars are also
present.  These C--stars cannot be disentangled using only a
photometric selection criterion (see also G04).

Figure \ref{fig:rebeirot}$c$ shows the histogram of the number of
photometrically selected C--stars versus $J-K_s$ colour, together
with the number of spectroscopically confirmed C--stars by RAW93
versus $J-K_s$.  There is a shift between the two distributions
suggesting that the spectroscopic identification of C--stars is
biased to bluer colours. This is not surprising -- in a sample
that was spectroscopically selected using CN bands near 8000 \AA,
Blanco et al.~(1980) also found more C--stars with bluer colours
than $J-K_s=2$.

We find that within the MACHO fields there are 117 stars with
$1.33\leq J-K_s\leq 2.0$ in the MC$^2$ that are not present in the
RAW93 catalogue. They amount to 17\% of the total number of MC$^2$
stars with these colours. Because of the bias discussed above this
is an upper limit to the number of O-rich stars contaminating this
region of CMD. From comparison with the spectroscopic sample of
G04, we also expect the contamination to be low due to the fact
that there are only 3 confirmed O-rich stars in the SMC with
$(J-K_s)_0=1.30$, $1.58$, and $2.80$, respectively.

Considering the mis--identifications or missed
cross--identifications as a result of an automatic association
criteria, we estimate a contamination of significantly less than
2.7\% and 7.9\%, respectively, based on Loup et al.~(2003).

\subsection{MACHO light--curves}
\label{subsection:machoselection}

\begin{figure}[h]
  \includegraphics[width=9cm]{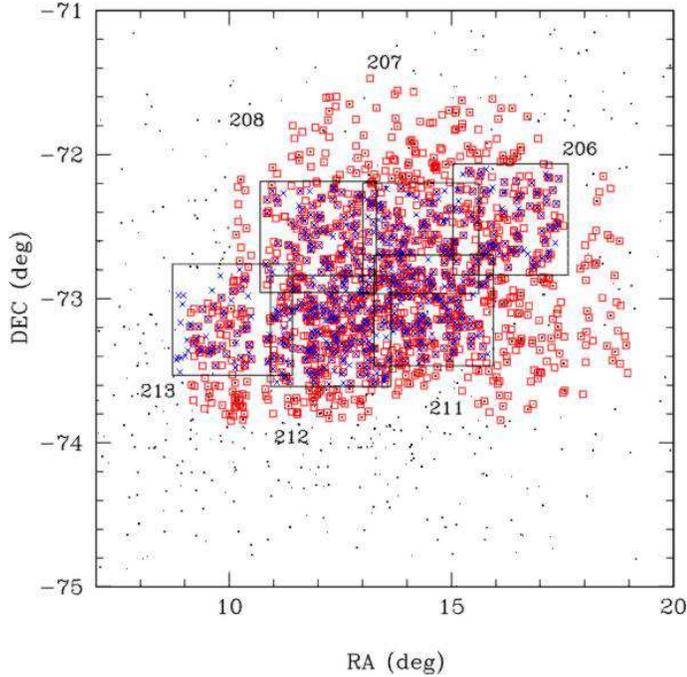}
   \caption{Locations of C--stars in the observed SMC MACHO fields
(6 large squares each indicated by its MACHO field number).
   Small black dots are all photometrically selected C--stars in the
   SMC; blue crosses refer to C--stars with $| d_{MACHO} - d_{2MASS} | \leq
   3$\arcsec.
   Small red squares are C--stars from the RAW93 catalogue cross--correlated
   with the $\mathrm{MC}^2$ catalogue.
   See the electronic edition of the Journal for a colour version
   of the figure.}  \label{fig:fields}
\end{figure}

\begin{figure}[h]
  \includegraphics[width=9cm]{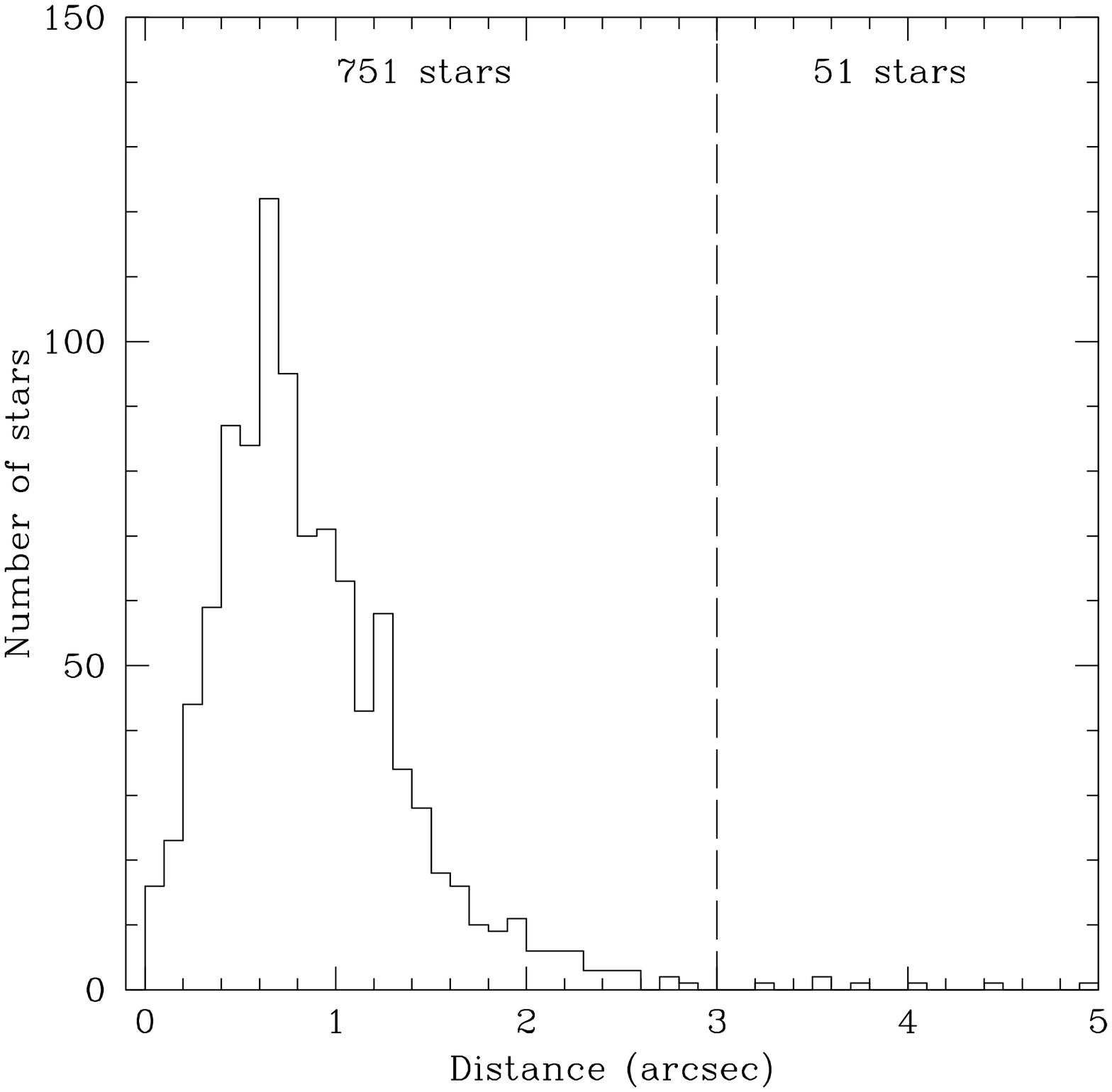}
   \caption{Histogram of the absolute distance between 2MASS sources and
    their MACHO counterparts. The dashed line is drawn at 3\arcsec.}
    \label{fig:hist_dist}
\end{figure}

The light--curves of stars in our sample were extracted from the
on-line MACHO catalogue\footnote{http://wwwmacho.mcmaster.ca/}.
The MACHO project observed the central body of the SMC
simultaneously in non--standard (i.e broad) blue ($B_M$) and red
($R_M$) bands for roughly 8 years (1992--2000). On average there
are 800--900 observations per filter for most stars.  Since we
were interested in the characteristics of the temporal behavior,
we used instrumental magnitudes and analyzed the photometric
variations in both bandpasses.

\begin{table*}[t]
\begin{center}
\caption{Cross-identification between the MASTER and MACHO catalogs of our sample of C-stars.}
\label{tab:table_id}
\small
\begin{tabular}{lcccccc}
\hline \hline
 MACHO       &     DCMC        &    2MASS      & $\alpha $    &    $\delta $    &  $d$ &  RAW93\\
\hline
213.15047.194  & J003516.05-732527.7 & 0035158-732527 &   8.816091 &  -73.424309  &  .100    &   0   \\
213.15048.4    & J003521.86-732422.8 & 0035216-732422 &   8.840220 &  -73.406326  & 2.400    &   0   \\
213.15054.264  & J003526.42-725835.6 & 0035263-725835 &   8.859856 &  -72.976501  & 1.150    &   0   \\
213.15046.550  & J003533.80-733032.6 & 0035337-733032 &   8.890656 &  -73.509033  &  .750    &   0   \\
213.15051.6    & J003537.30-730956.4 & 0035372-730956 &   8.905227 &  -73.165588  &  .840    &   0   \\
213.15048.8    & J003538.53-732441.3 & 0035384-732441 &    8.910099 &  -73.411438 &  1.910  &     0  \\
213.15053.2   &  J003547.99-730213.7 & 0035479-730213 &    8.949644 &  -73.037003 &   .430 &      0  \\
213.15054.103 &  J003552.13-725834.2 & 0035520-725834 &    8.966831 &  -72.976120 &   .650 &      0  \\
213.15105.8   &  J003603.01-732346.3 & 0036029-732345 &    9.012422 &  -73.396103 &  1.210 &      0  \\
213.15106.14  &  J003612.60-731711.7 & 0036125-731711 &    9.052299 &  -73.286476 &   .360 &      0  \\
213.15105.12  &  J003616.80-732133.1 & 0036167-732133  & 9.069947   & -73.359169    & .660  & 0   \\
213.15108.5   & J003628.84-731144.1 & 0036287-731144 & 9.119937 & -73.195580  & .170  &    1 \\
213.15108.7   & J003630.13-731033.8 & 0036300-731033  &  9.125388  & -73.175995  &  .380  & 0 \\
213.15109.14  & J003633.23-730548.0 & 0036331-730547   & 9.138298 & -73.096611 &   .320    &   3 \\
213.15106.12  & J003648.03-731830.7 & 0036479-731830  &  9.199909  & -73.308487 &  .320 & 6 \\
\hline \hline
\end{tabular}
\end{center}
\end{table*}

2MASS and MACHO coordinates were cross-correlated using a search
radius of 3\arcsec\ and the nearest, and reddest star from MACHO
catalogue was chosen as the counterpart to the 2MASS source. Of
the 802 photometrically selected C--stars lying  within the MACHO
fields (Fig.~\ref{fig:fields}), 25 stars were detected twice with
different identification numbers because of overlap between
adjacent MACHO fields. However, since we chose the nearest star to
a 2MASS star as a MACHO counterpart, we avoid double or multiple
identifications of the same variable. Nevertheless we did check
that the quality of the light--curves is similar. The histogram in
Fig.~\ref{fig:hist_dist} shows the distance in arcsec between a
2MASS source and the corresponding star in the MACHO database. The
distribution peaks at $d\sim 0$\farcs6, and 65\% of the stars are
within $d\leq$1\arcsec. For 51 stars the nearest MACHO counterpart
is at $d>3$\arcsec. Of these, 20 are within 10\arcsec (6 within
4\arcsec) and 31 within 20\arcsec. The majority of these stars are
located at the edges of the MACHO fields and have poorer
astrometry perhaps due to distortions. Moreover, they have bluer
instrumental magnitudes than other stars in our sample. Therefore
they have been excluded from our analysis.  Finally, the
photometrically selected sample contains 751 stars with MACHO
light--curves.

By cross-correlating the RAW93 and $\mathrm{MC}^2$ sample with the
MACHO data set, we found that 328 spectroscopically confirmed
C--stars are not included in our photometrically selected sample.
Thus, the total number of C--star light--curves analysed in the
next section is 1079 (751+328).

Table~\ref{tab:table_id} is an extract of the full table,
available electronically at Centre de Données astronomiques de
Strasbourg (CDS)\footnote{http://cdsweb.u-strasbg.fr/}, and
reports the first 15 lines of the cross-identified MACHO and
$\mathrm{MC}^2$ sources.  It contains: MACHO, DCMC, and 2MASS
identifier (Cols.\ 1-3); right ascension and declination (in
degrees) from the second incremental release of the 2MASS
catalogue (Cols.\ 4 and 5); the positional difference between
2MASS and MACHO coordinates in arcsec (Col.\ 6) and the RAW93
identification number (if appropriate, otherwise 0) (Col.\ 7).

\section{Analysis of Periods and Amplitudes}
\label{section:analysis}

An independent  Fourier analysis of the $B_M$  and $R_M$
light--curves was  performed to  search for  periodicities in  the
data.   The MACHO time-baseline is  about 2700  days, more than
twice that  of OGLE--II database used, for example, by Ita  et
al.\ (2004a, 2004b), Kiss \& Bedding  (2004) and G04. Resulting
periods   above  2600  days ($\log P = 3.4$) should  be considered
less  significant.   Only  photometric  measurements that are more
accurate than  0.1  mag  are  used.  The  method  that extracts
the light--curve  parameters  is   based  on  the Lomb-Scargle
algorithm \citep{Lomb76,Scargle82} as used by Rejkuba et al.\
(2003). It is described in Appendix A.1.

Initially 751 light--curves of photometrically selected C-stars
were analysed. All the light--curve fits and their parameters were
inspected visually. Based on this inspection, and on parameters
returned by the light--curve analysis programmes, three quality
flags were assigned to each light--curve:  $flag(1)$ describes the
data quality, $flag(2)$ describes the light--curve fit quality,
and $flag(3)$ describes the detected periodicities. Table
\ref{tab:table_flags} summarizes flag values and their meaning
while more details are given in Appendix A.2 with examples of
light-curves associated to a given flag value. Light--curves with
$flag(1) \leq 2$ produce reliable period determinations, as
confirmed by the fact that for most of them the Fourier analysis
has provided good results ($flag(2)\leq2$). Only for some stars
($8\%$ and $11\%$, respectively, in $B_M$   and  $R_M$ photometry)
classified as good light curves ($flag(1)\leq2$), uncertain
periodicity ($flag(2)=3$) or no periodicity ($flag(3)=0$) was
detected because of highly irregular light--curves.

\begin{table}
\caption{Description of the values of different $flags$ for the complete sample of 1079 C--stars. }
\label{tab:table_flags}
\begin{center}
\begin{tabular}{lllll}
\hline \hline
\multicolumn {3} {c} {Data Quality} & BLUE & RED \\
\hline
$flag(1)$ & 0 & excellent              & 306 & 803 \\
          & 1 & good                   & 478 & 70 \\
          & 2 & fair                   & 136 & 20 \\
          & 3 & noisy or few data      & 115 & 150 \\
          & 4 & no data                & 44  & 36 \\
\hline
\multicolumn {3} {c} {Fit Quality}     & BLUE & RED \\
\hline
$flag(2)$ & 0 & excellent                  & 124 & 101 \\
          & 1 & good                       & 191 & 258 \\
          & 2 & fair                       & 470 & 349 \\
          & 3 & bad                        & 127 & 168 \\
          & 4 & underivable                & 123 & 167  \\
          & 5 & underivable: $flag(1)$=4   & 44  & 36  \\
\hline
\multicolumn {3} {c} {Detected Periodicity} & BLUE & RED \\
\hline
$flag(3)$ & 0 & no periodicity             & 143 & 135 \\
          & 1 & 1 period                   & 558 & 608 \\
          & 2 & 2 periods                  & 311 & 292 \\
          & 3 & one period                 & 67  & 44  \\
          &   & second period uncertain    &     &     \\
\hline \hline
\end{tabular}
\end{center}
\end{table}

The same procedure was applied to the 328 stars common to MC$^2$,
RAW93 and MACHO but missed by the photometric selection. Aliases
were identified  from the diagram $log\, P$ $vs.$ $K_s$ magnitude
as those periods that create clear vertical paths (see also
Appendix A.1). These correspond to periods equal to  1 and  2
years exactly and were removed from our analysis.

The adopted procedure allowed us to define a semi-automatic
algorithm to obtain the light--curve parameters and access their
quality. It is summarized as follows: the best fitting period(s)
are obtained from Eq.~\ref{secondperiod};  the quality of  the
observations is derived from  $flag(1)$; the quality of the period
determination is evaluated using the spectral power that is
closely related with a semi-automatic definition of $flag(2)$
(Appendix A.2).  Note that this procedure can also be applied to
stars of a different type (i.e. M--type stars) in the MACHO
catalogue or to other measurements of stellar variability in a
comparable sampling.

Amplitudes related to the main periodicity of light--curve
variations were determined both from the sinusoidal fit of each
light-curve and from the peak-to-peak magnitude difference. A
comparison of both determinations is given in Appenxix A.3.

\begin{figure}
\center \includegraphics[width=9cm]{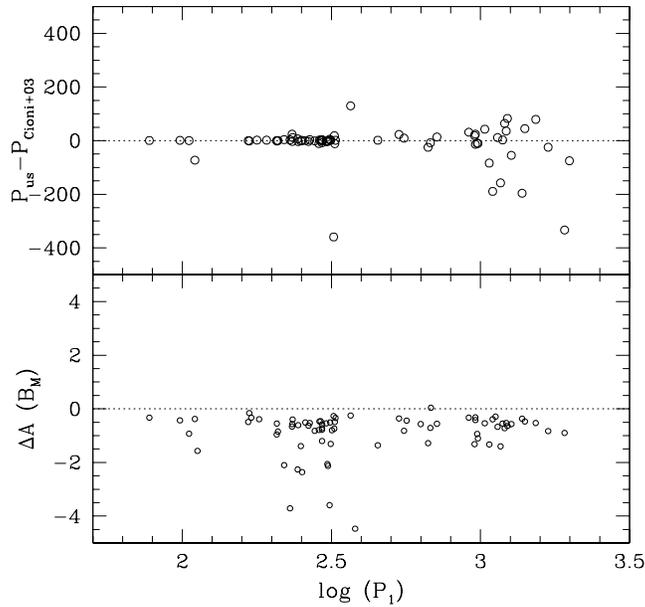}
   \caption{Comparison between the period and amplitude derived in this work
     and those from Cioni et al.\ (2003).}
   \label{fig:cioni03}
\end{figure}
\begin{figure}
\center \includegraphics[width=9cm]{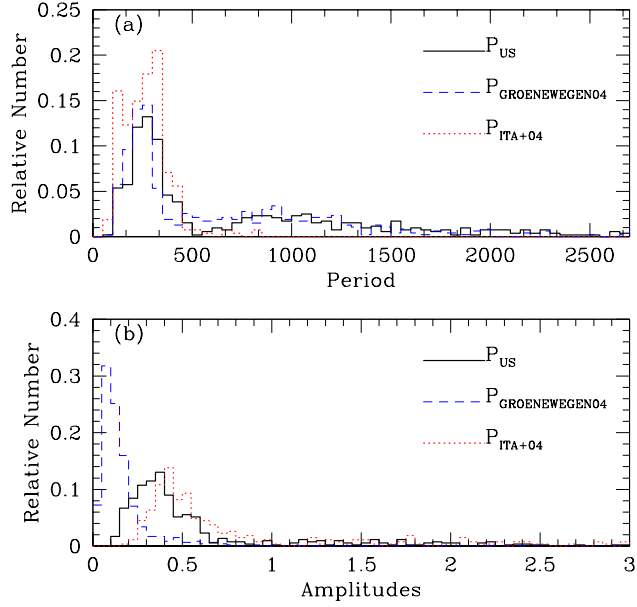}
   \caption{Comparison between the period and amplitude derived in this work
     and those from Ita et al.\ (2004b) and G04.}
   \label{fig:Ita+04&G04}
\end{figure}

\subsection{Results and statistics}
\label{section:statistic}

\begin{table*}[t]
\begin{center}
\caption{Light curves parameters: $X$ is the mean magnitudes; first (second) row refers to $B_M$($R_M$)-light curve.}
\label{tab:table_BR}
\tiny
\begin{tabular}{rrrrrrrrrrrccc}
\hline \hline
 MACHO       &
$<X>$        &     $A_1$       &    $B_1$      &
$A_2$        &     $B_2$       &    $ P_1$     &
$P_2$        &     ${\cal P}_1$ &    ${\cal P}_2$ &
$A_{phot}$   &
$flag(1)$    &     $flag(2)$    &     $flag(3)$  \\
\hline
213.15047.194  &         .000 &       .000 &       .000 &       .000 &       .000 &       .000 &       .000 &       .000 &       .000 &      .00&       3 &      4 &      0 \\
 &                       .000 &       .000 &       .000 &       .000 &       .000 &       .000 &       .000 &       .000 &       .000 &      .00&        2 &      4 &      0 \\
213.15048.4    &       -8.139 &       .020 &      -.077 &       .027 &      -.020 &   1002.284 &    199.586 &    102.100 &     46.200 &      .26&       0 &      1 &      2 \\
 &                     -9.629 &       .060 &       .011 &       .000 &       .000 &   1658.224 &       .000 &     83.200 &       .000 &      .16 &        0 &      2 &      1 \\
213.15054.264  &         .000 &       .000 &       .000 &       .000 &       .000 &       .000 &       .000 &       .000 &       .000 &      .00&       4 &      5 &      0 \\
 &                     -5.808 &      -.743 &      -.917 &      -.300 &       .776 &   1696.516 &    359.045 &    178.700 &     96.400 &     1.91 &        1 &      0 &      2 \\
213.15046.550  &         .000 &       .000 &       .000 &       .000 &       .000 &       .000 &       .000 &       .000 &       .000 &      .00&       4 &      5 &      0 \\
 &                       .000 &       .000 &       .000 &       .000 &       .000 &       .000 &       .000 &       .000 &       .000 &      .00 &        3 &      4 &      0 \\
213.15051.6    &       -7.740 &      1.051 &       .479 &       .000 &       .000 &    621.195 &       .000 &    319.800 &       .000 &     2.48&       1 &      1 &      1 \\
 &                     -9.615 &       .589 &       .598 &       .000 &       .000 &    611.472 &       .000 &    235.600 &       .000 &     1.59 &        0 &      1 &      1 \\
213.15048.8    &       -7.242 &       .154 &       .025 &       .000 &       .000 &    171.377 &       .000 &    157.200 &       .000 &      .50&       1 &      1 &      1 \\
 &                     -8.822 &       .104 &       .062 &       .000 &       .000 &    171.699 &       .000 &    114.900 &       .000 &      .30 &        0 &      1 &      1 \\
213.15053.2    &       -6.844 &       .195 &       .133 &       .069 &      -.111 &    857.187 &    230.560 &    221.400 &     65.800 &      .57&       1 &      1 &      3 \\
 &                     -8.804 &       .153 &       .093 &       .061 &      -.057 &    870.538 &    230.751 &    248.100 &     69.100 &      .44 &        0 &      1 &      2 \\
213.15054.103  &         .000 &       .000 &       .000 &       .000 &       .000 &       .000 &       .000 &       .000 &       .000 &      .00&       4 &      5 &      0 \\
 &                     -6.303 &       .261 &      1.009 &       .103 &      -.953 &    406.026 &   2287.050 &    123.900 &     97.300 &     1.73 &        2 &      0 &      2 \\
213.15105.8    &       -7.505 &      -.417 &      -.014 &       .000 &       .000 &    876.287 &       .000 &    307.400 &       .000 &     1.11&       1 &      1 &      1 \\
 &                     -9.279 &      -.339 &      -.027 &       .000 &       .000 &    873.740 &       .000 &    220.800 &       .000 &      .61 &        0 &      0 &      1 \\
213.15106.14   &       -7.132 &      -.232 &       .085 &       .147 &       .010 &    294.109 &    151.497 &    163.500 &     51.300 &      .59&       1 &      1 &      3 \\
 &                     -9.088 &       .005 &      -.138 &      -.108 &       .017 &    260.586 &    300.278 &     80.000 &     69.800 &      .32 &        0 &      2 &      3 \\
\hline \hline
\end{tabular}
\end{center}
\end{table*}

Table~\ref{tab:table_BR}  lists   the  parameters derived from our
analysis   for  a  total   of  1079 stars. Only the first ten
lines are shown in  this paper, but the complete table is
accessible electronically  via CDS.  The table contains: MACHO
identifier (Col.  1);   the  quantities of the period--amplitude
analysis for $B_M$ (first  row) and $R_M$ (second row) light
curves: (Cols. 2-9):  mean magnitude $X_0$, $A_1$, $B_1$, $A_2$,
$B_2$, $ P_1$, and $P_2$ in days, power strength of  the first
${\cal P}_1$  and second  period ${\cal P}_2$;  $A_{phot}$   (Col.
10)  $flag(1)$, $flag(2)$,  and $flag(3)$ values  (Col. 11-13).
Among the 1079 stars the following have passed the data--quality
criterium $flag(1)\leq 2$: 919 $B_M$ and 893 $R_M$ light-curves.

In the following discussion, only the good 919 $B_M$ light-curves
are used, unless explicitly stated otherwise. Within this sample
785 stars also have $flag(2)\leq 2$ and a minimum amplitude of
0.05 mag, thus are all variables. About 4\% show a very regular
variation with only one periodicity. The others appear
multiperiodic. Some (7\%) show a well--defined first period with
amplitude variations and a less clear second periodicity, while
others (59\%) clearly show two periods (examples are given in
Fig.~\ref{fig:lc_flag3}).

The  availability of accurate  photometry and  long--term
observations has made the distinction between regular and
semi--regular (SR) variables more and more difficult (Whitelock et
al.~1997). The stellar light--curves  can be  as regular in  SR as
in Mira class, but on average  SR variables show smaller
amplitudes  (Cioni et al.\ 2003). However,  due to  a difficult
and rather subjective classification  we only distinguish between
two broad groups: \emph{sources which show  a clear single
periodicity} and \emph{sources which are multiperiodic}. About
15\% appear to be irregulars ($flag(2)=3$), with no clear period.

Table~\ref{tab:table_BR} has 131 stars are  in  common with Cioni
et al.~(2003). Fig.~\ref{fig:cioni03}$a$  shows the comparison
between the periods derived  in  the present paper and those in
Cioni et al.~(2003). Amplitudes are compared in
Fig.~\ref{fig:cioni03}$b$. We plotted stars  with $flag(2)\leq2$
in our  analysis and  stars with $Flag   <  9$   and  $Flag   \ne
5$ in  the   table  by Cioni et al.~(2003). There are 78 stars in
common. The mean period and amplitude differences are:
$P_{present}-P_{Cioni03}=-14 \pm   76$ and
$A_{present}-A_{Cioni03} =-0.9  \pm  0.8$. The periods agree
within the uncertainty in the period determination that is on the
order of 5\%, except for few stars for which Cioni et al.\ (2003)
derive only one period and flag the LPV as multiperiodic, while we
find 2 periods that are typically shorter. It is possible that
additional longer periods are present as well. In contrast,
amplitudes are systematically different. Note that Cioni et al.
(2003) define amplitude as the difference between the minimum and
 maximum value of MACHO photometry, which is different from our
definition of $A_{phot}$ (see Appendix A.3). This could explain
why Cioni et al.\ amplitudes are systematically larger than ours.

Figure~\ref{fig:Ita+04&G04}  shows the comparison between periods
and amplitudes as derived in the present paper and those by Ita et
al.~(2004b) and G04. In both panels, for all the three works only
stars satisfying our photometric criteria are reported. In
Fig.~\ref{fig:Ita+04&G04}$a$ in the case of multi--periodic
variables the period of G04 corresponding to the largest amplitude
and our first period are reported. Ita et al. (2004b) only give
the predominant period and do not analyze multi--periodic
light--curves. The present results are in good agreement with
those by G04, predicting periods as long as 2400 d, while Ita et
al. (2004b) do not find periods longer than about 690 d.

In Fig.~\ref{fig:Ita+04&G04}$b$ the OGLE $I$--band amplitudes of
Ita et al. (2004b) and G04 are compared with the present
$R_M$--band amplitudes as derived from the light--curves
($A_{phot}(R_M)$, see Appendix A.1). In the case of G04 we plot
the largest amplitudes from his Table 2. Although a direct
comparison between $R_M$-- and $I$--band amplitudes is difficult,
the shape of the histogram illustrating our results is similar to
that by Ita et al. (2004b), even if we have a larger number of
stars with amplitudes smaller than $\sim$0.4. G04 found that the
majority of C--stars in his sample have $\Delta I < 0.2$, while
Ita et al. (2004b) have almost no stars in the same amplitude
range.

Both Fig.~\ref{fig:cioni03}$b$ and Fig.~\ref{fig:Ita+04&G04}$b$
indicate that the various methods for estimating pulsation
amplitudes can give systematically different results (see also
Appedix A.3). Ita et al. (2004b) and Cioni et al. (2003) derive
pulsation amplitudes as $\Delta  X = X_{max} - X_{min}$, where $X$
is the photometric band used, while G04 estimate amplitudes from
the sinusoidal fit.

\begin{figure}
\center
\includegraphics[width=9cm]{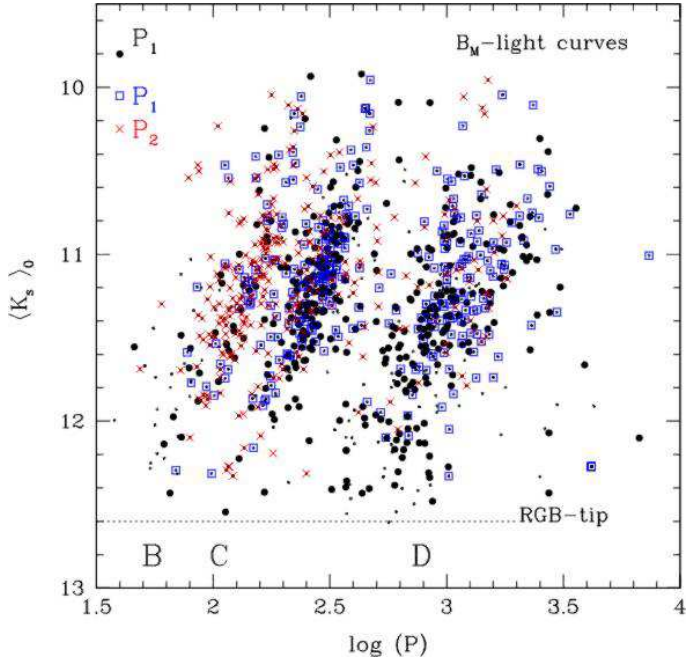}
   \caption {$\langle K_s \rangle _0$ $vs.$ $logP$ relation
    for all C--stars in the sample
   (small black dots). Large dots are first periods ($P_1$) for
   stars with $flag(2)\le 2$ and $flag(3)=1$; blue open squares $P_1$
   for stars with $flag(2)\le2$ and $flag(3)=2$ and 3; red
   crosses correspond to the second period ($P_2$) of stars indicated by
   squares. The $\langle K_s \rangle _0 $ magnitude is the average of DENIS
   and 2MASS measurements, dereddened as explained in the text.
   Letters identify the parallel sequences as
   defined by Wood et al\ (1999).}  \label{fig:logPK}
\end{figure}
\begin{figure}
\center
\includegraphics[width=9cm]{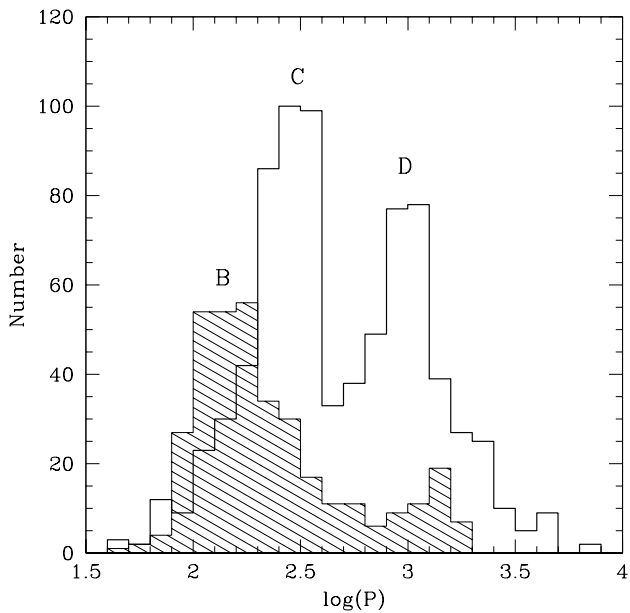}
   \caption{Histogram of all the periods found for $B_M$ light--curves,
though for $R_M$ light--curves results are very similar. The empty
histogram shows the distribution of P$_1$ and the shaded histogram
of P$_2$. Letters indicate sequences as in Wood et al. (1999).}
   \label{fig:histoperiods}
\end{figure}

\section{Discussion}
\label{section:discussion}
\subsection{Log(P) vs. $K$ diagram}

In the past five years new results for the log(P) vs. $K$--band
mag diagram have been obtained as a by--product of microlensing
projects such as MACHO, OGLE, and EROS. Wood et al.\ (1999) found
that the bright red giant variables in the LMC form four
sequences: three are the result of different pulsation modes and
one, at the longest periods (seq. D or the long secondary period
sequence -- LSP), remains unexplained. Ita et al. (2004b) showed
that some sequences possibly split into sub--sequences at the
discontinuity around the tip of the RGB. G04 analyzed
spectroscopically confirmed M-- and C--type stars and concluded
that the LSP sequence is independent of evolutionary and chemical
(C--rich or O--rich) effects. A comparison of the SMC and the LMC
sequences can also be found in Cioni (2003). More recently,
Schultheis, Glass, and Cioni (2004) compared variable stars in the
different sequences between the Magellanic Clouds and the NGC6522
field in the Galaxy. These authors conclude that all three fields
contain similar types of variables, but the proportion of stars
that vary decreases at lower metallicities and the minimum period
associated with a given amplitude gets longer.

Figure \ref{fig:logPK} shows the distribution of the mean $\langle
K_s \rangle _0$ mag vs.\ the logarithm of the period obtained by
analyzing $B_M$ light--curves. However, the discussion that
follows applies also to a similar analysis of $R_M$ light--curves.
In fact as expected from the comparison shown in
Fig.~\ref{fig:PBR}, there are no differences in the periods
obtained from the two channels.

In Fig.~\ref{fig:logPK} all sources in the sample are indicated by
small black dots. In addition, stars for which only one reliable
period was detected ($flag(2)\le2$ and $flag(3)=1$) are plotted as
larger dots, while for stars with two reliable periodicities
detected ($flag(2)\le2$ and $flag(3)=2$ and 3), the first and the
second periods are plotted with blue open squares and red crosses,
respectively. The $K_s$--band photometry is the mean of DCMC and
2MASS measurements ($\langle K_s \rangle$). DCMC magnitudes are
corrected for the shift in the absolute calibration according to
Delmotte et al.\ (2002). The $K_s$--magnitude is also dereddened
according to a SMC mean reddening of $E(B-V)=0.065\pm0.05$
obtained by averaging different measurements \citep{Westerlund97}.
By adopting the extinction law of Glass (1999) and $R_V=3.1$ we
obtain $A_{K_s}=0.02$ and $A_J=0.05$.

All periods range in the interval 1.5$\lsim $ log(P)$\lsim $ 3.5
(Fig.~\ref{fig:logPK}). There are 3 well-defined parallel
sequences in the PL diagram with a small number of stars lying
between the sequences. Part of the scatter is due to the fact that
the $K_s$--band is a mean of only 2 measurements and part is due,
probably, to depth effects in the SMC. The line of sight depth of
the SMC is estimated to range between 5--20 kpc by various authors
\citep{Westerlund97}. Hence, assuming the SMC distance modulus of
(m-M)=19~mag and the full depth of 10 kpc, we expect a scatter on
the order of $\sim 0.4$~mag around the average.

Sequences can be identified with B, C, and D from Wood et al.
(1999). It is interesting to note that C-stars do not populate
shorter sequences (i.e. sequence A), a result already visible from
Fig.~4 of Ita et al.~(2004b), and noted in the LMC by Fraser et
al.~(2005). Since the baseline explored here is longer than that
of previous works (except for Fraser et al.~2005 who analysed
MACHO data in the LMC), we find a well populated D sequence, about
34\% of the variable stars in our sample have a first or second
period longer than 630 days. This should be compared with previous
results which found ($i$) 25\% of all variable AGB stars in the
LMC \citep{Wood+99},  but with a very small fraction belonging to
C-rich LPVs (Fraser et al. 2005), and ($ii$) 24.6\% of all the
spectroscopically selected C-stars with periods from OGLE-II
photometry in the SMC on the sequence D (G04). The bias against
detection of variables with periods in excess of $\sim 800$ days
in the latter work may be why we find more C-stars on this
sequence. This shows that very long--term monitoring is essential.
Even in the case of our sample from MACHO with an 8-year
time--baseline, it is not clear if the drop in the period
distribution at $\log\,P \sim 3.3$ is real or an artifact due to
incompleteness at longest periods (see
Fig.~\ref{fig:histoperiods}).

Sequence D is much broader than the others, and the nature of its
stars is still a matter of debate \cite{wood+04}. Only a few of
them, those with \mbox{$J-K_s>2$} (see Sect.~\ref{sect:BCD_CMD})
are probably dust-enshrouded AGB stars that could have either
carbon or oxygen-dominated chemistry, but their number is very
small. Thus, as already concluded by Wood et al.\ (2004) from the
similarity of the colour variations associated with both the
primary and secondary periods, dust is unlikely to cause the LSPs.
Clearly, a long-term spectroscopic and photometric monitoring of
these stars is necessary to gain some insight into the nature of
their variability.

In Fig.~\ref{fig:histoperiods} we compare the histogram of the
first detected periods, which represent the dominant periodicity
in a given star, with the second best period. First periods mainly
occupy sequences $C$ and $D$ with the peaks at $log\,P=2.45$ and
$log\,P=2.95$, respectively. Second periods mainly populate
sequence $B$ and peak at $log\,P =2-2.3$. A weak peak is also seen
at $log\,P= 3.2$, which is clearly part of sequence $D$.

\subsection{Log(P) vs. period ratio}

According to Lattanzio  \& Wood (2003) a star that  evolves up the
AGB first  pulsates at  low  amplitude  on sequence  A,  then with
further evolution the sequence  B mode will become unstable  and
its amplitude increases, while  the sequence A mode amplitude
decreases. The star will be a multi--mode pulsator having periods
in sequences $A$ and $B$.  The  pulsation amplitude of each mode
increases with increasing stellar radius.  Subsequently a
fundamental mode in  sequence $C$ will also become unstable. At
this  moment up to three different periods of pulsation  can be
detected in  a given  star. However,  the competing growth  rate
of the  amplitude of  each pulsating  mode may  shade the
detectability of a given periodicity.  For example for a $1
M_{\odot}$ star  the  fundamental period  dominates  at
$M_{bol}>-4.5$, while  at $M_{bol}<-3.5$  the  first and  second
overtone  pulsation modes  are stronger. The  star will finally
end up as a  dust--enshrouded, large amplitude  fundamental mode
pulsator prior to ejection  of  all its envelope and the beginning
of the Planetary Nebulae phase.

\begin{figure}
\center
\includegraphics[width=9cm]{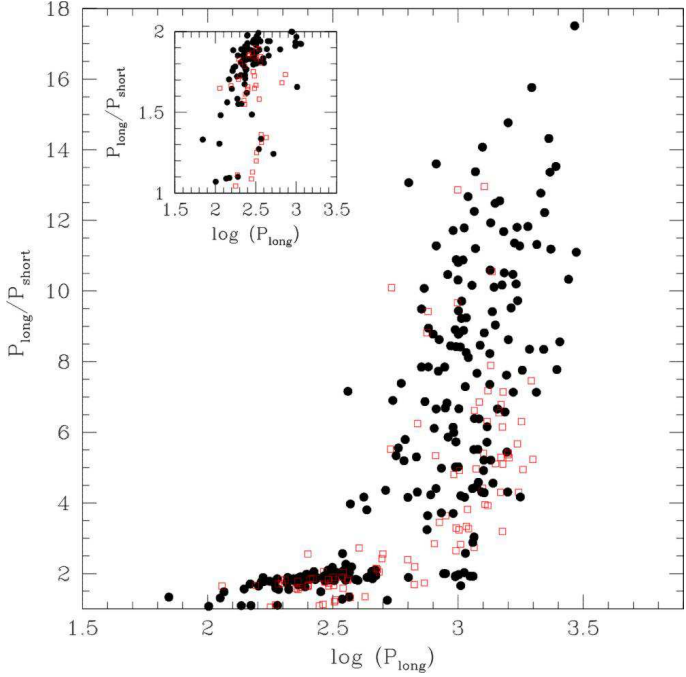}
   \caption{The ratio between the longer to the shorter period versus the
    longer period.
    Black--filled circles represent stars with $P_1>P_2$, while red open
    squares sources with $P_1<P_2$. In the upper left corner we show an
    enlargement of the lowest ratios.}
  \label{fig:logPratio}
\end{figure}

For    stars    with   detected    multiperiodicity    we   plot
the  ratio between the longer  to the shorter period as a function
of the longer one in Fig.~\ref{fig:logPratio}. A well--defined
group of stars have ratios  ranging from 1  to 2. These are shown
in zoom--in  in the upper  left corner  of the  figure. They
belong to  the  sequence $C$ (longer)  and $B$  (shorter) periods.
Their ratio  distribution agrees with  the scenario proposed  by
Lattanzio  \& Wood  (2003) for  a star pulsating  in the
fundamental,  first and  second overtone (see their Fig.~53). An
extended  tail up to a ratio of 20  is populated by stars with the
longer period on sequence $D$ and the shorter period along either
sequence  $C$ or  $B$. There are no  theoretical models available
at present  to  explain these high  ratios  which involve an
intrinsic stellar  pulsation;   thus a  different nature for
long--term modulation needs to be invoked. It is interesting to
note that most of the sources  with $P_1<P_2$ have lower period
ratios and slightly larger values of longer periods with respect
to sources with $P_1>P_2$.

\subsection{Log(P) vs. amplitude}
\label{sect:amplitude}

\begin{figure}
\center
\includegraphics[width=9cm]{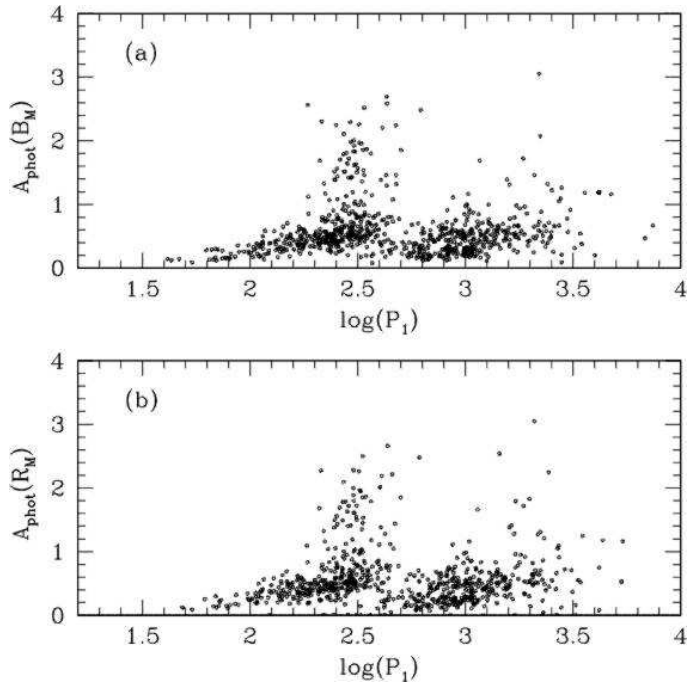}
   \caption{Period distribution as a function of amplitude in
the two MACHO photometric bands.} \label{fig:BRJKampl}
\end{figure}

Figure \ref{fig:BRJKampl} shows the period distribution (of only
the first periods $P_1$) as a function of amplitude in the MACHO
bands. Most variables have amplitudes in the optical bands below
about $0.5$ mag. Stars that occupy sequences $C$ and $D$ are
clearly separated. Sequence $B$ is not present in these diagrams
because it is mostly populated by secondary periods. It should be
also noted that amplitudes belonging to secondary periods are
typically smaller, and thus this figure should be compared with
those of other authors (e.g.\ G04, Ita et al.\ 2004b) with
caution. In addition, the definition of amplitude is not always a
trivial issue with these highly variable stars that often have
variable amplitudes as well. We discuss this in Appendix A.3. Here
we use amplitudes determined directly from photometry ($A_{phot}$)
as defined in Appendix A.3.

Both  $C$ and $D$ sequences present distribution tails to large
amplitude (from a few tenth up to 3 mag) values that probably
correspond to variables of Mira type. While Miras typically occupy
sequence $C$, a few of them can be found on sequence $D$ because
they have fainter magnitudes due to dust obscuration.

\begin{figure}
\center
\includegraphics[width=9cm]{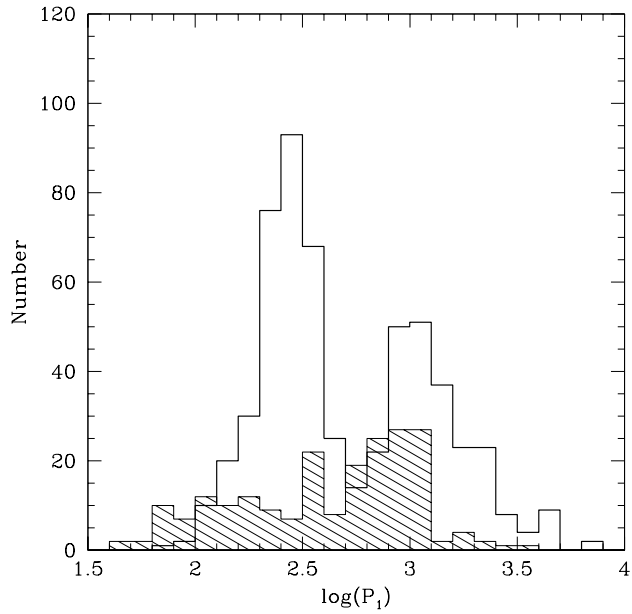}
   \caption{Period distribution ($log(P_1)$) of stars with
    $0.05\leq A_{phot}(B_M)\leq 0.2$ (empty) and $A_{phot}(B_M)> 0.2$
    (shaded).}
   \label{fig:histlogPamp}
\end{figure}

Figure \ref{fig:histlogPamp} shows the period distribution for small
($0.05\leq A_{phot}(B_M)\leq 0.2$; open histogram) and large
($A_{phot}(B_M)> 0.2$; shaded histogram)
amplitude variables. The two distributions are different. Small
amplitude variables have two peaks that correspond to the period--magnitude
relations $C$ and $D$. Large amplitude variables are more homogeneously
distributed between $log(P_1)=1.6$ and $3.1$. Fewer of them have
longer periods, though at $\log\, P \ga 3.1$ the sample might be incomplete.

In Schultheis, Glass \& Cioni (2004) the period of both small and
large amplitude variables defined as in Fig.~\ref{fig:histlogPamp}
increases progressively with decreasing metallicity, even though
the general period distribution of the two classes is fairly
similar. In the present paper stars located in sequence $C$
($1.9\leq log(P_1)\leq 2.7$) span the full range of periods if
they are either small or large amplitude variables. Therefore
there seems to be no indication of a difference in metallicity
between the small and large amplitude LPVs on this sequence. On
the other hand, the majority of large amplitude variables that
occupy sequence $D$ ($2.7\leq log(P_1)\leq 3.5$) have
$log(P_1)\leq 3.1$. A longer monitoring time--baseline is
necessary to discern whether this lack of longer period large
amplitude LPVs on sequence $D$ is real or due to incompleteness,
and thus if it could represent shortage of lower metallicity stars
among the large amplitude variables.


\subsection{Log(P) vs. $J-K_s$ colour}

\begin{figure}
\center
\includegraphics[width=9cm]{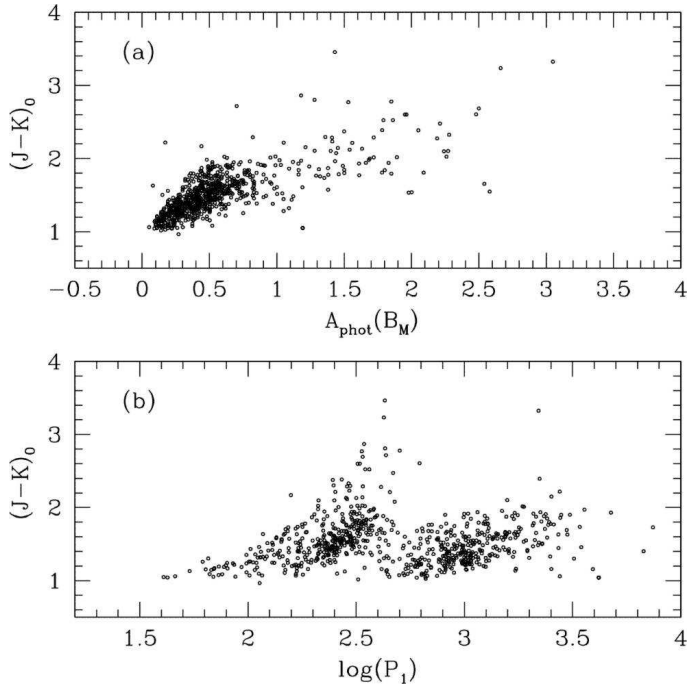}
   \caption{$(J-K_s)_0$ vs. the $B_M$-band amplitude $(a)$ and vs. $log\,(P_1)$
   $(b)$ for all C--stars.}
   \label{fig:BJKcol}
\end{figure}

Figure \ref{fig:BJKcol} shows that redder variables have a larger
amplitude. In fact Ita et al.~(2004b) also found that the C--rich
regular pulsators (Miras) have larger amplitude the redder the
star, while the O--rich Miras seem to have arbitrary amplitudes as
a function of $J-K_s$ colour. When the $J-K_s$ colour gets redder,
periods increase for variables in the $C$ sequence (see Figure
\ref{fig:BJKcol}$b$). This is apparently not the case for stars
populating the $D$ sequence, which again indicates a different
mechanism is responsible for the light--curve variations.

\subsection{Properties of stars on the $BCD$ sequences}
\label{sect:BCD_CMD}

\begin{figure}
\center
\includegraphics[width=9cm]{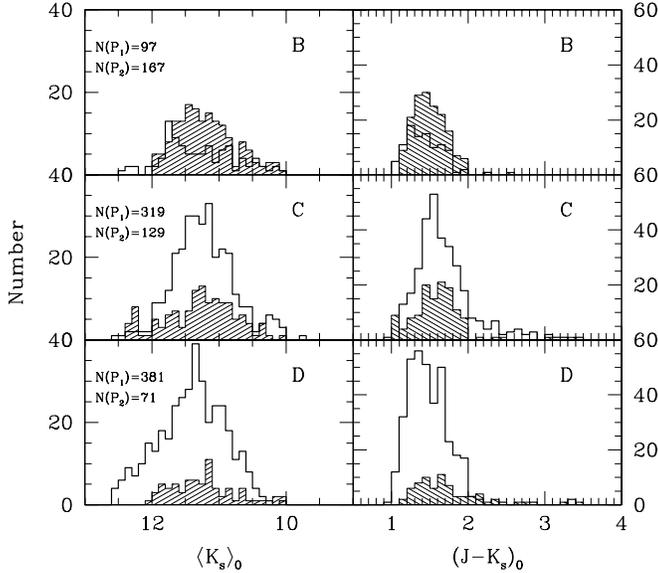}
   \caption{Distribution of the number of sources populating each $BCD$
   sequence versus $K_s$ magnitude and versus $J-K_s$ colour. The empty
   histogram indicates the first period ($P_1$) and the shaded histogram
   the second period ($P_2$). }
   \label{fig:histseq}
\end{figure}

We found no difference in the location of stars belonging to the
$BCD$ sequences in the $K_s$ vs. $J-K_s$ diagram. In Fig.
\ref{fig:histseq} we investigated the $K_s$ luminosity function
and $J-K_s$ colour distribution of stars in each sequence for both
the first (empty histogram) and second (shaded histogram) detected
periods. Most of the stars with $J-K_s>2$ belong to sequence $C$,
and they have only one detected and dominant periodicity. A
handful of them show a long second period in sequence $D$. The
faintest stars analyzed in this sample ($12\leq K_s \leq 12.6$)
have their first period either in sequence $B$ or $D$ and
eventually a second period in sequence $C$, which is often the
case for stars in sequence $D$ that are multi--periodic, while
stars in sequence $B$ usually have only one period detected. The
bulk of the C--star population has a first period that occupies
sequence $C$ or $D$ and somewhat less sequence $B$, which is
instead largely populated by the second period of these same
stars.

\subsection{Log(P) vs. $M_{bol}$}

\begin{figure}
\center
\includegraphics[width=9cm]{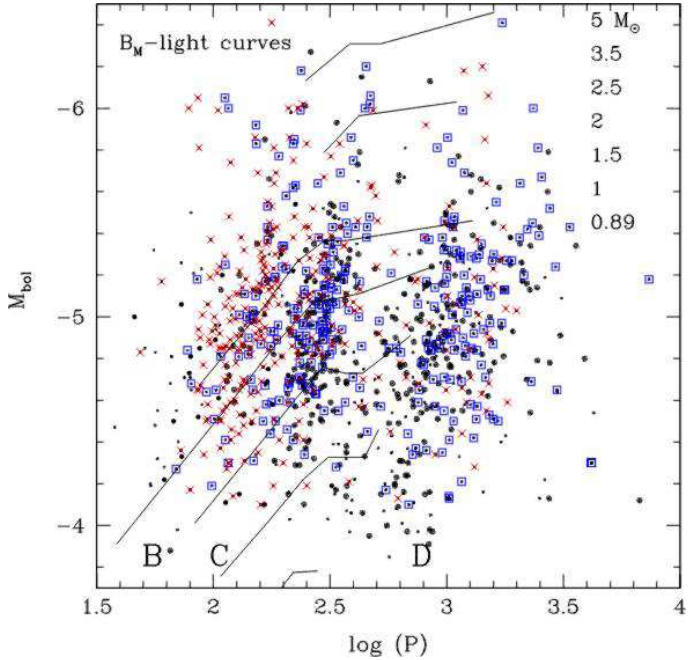}
   \caption{$M_{bol}$ $vs.$ $logP$ relation for all stars in the sample
   and overplotted theoretical tracks by Vassiliadis \& Wood
   (1993). Green triangles indicate stars with $(J-K_s)_0 > 2.1$
   and $flag(2)\leq 2$, while other symbols are as in Fig.~\ref{fig:logPK}.
   Masses are indicated on the right hand side of each track.
   See the
   electronic edition of the Journal for a colour version of the
   figure.}  \label{fig:logPMbol}
\end{figure}

Figure \ref{fig:logPMbol} shows the PL--relation for all stars in
the sample. Overplotted are the theoretical models by Vassiliadis
\& Wood (1993) for different masses and a mean SMC metallicity of
$Z=0.004$. To transform $K_s$ magnitudes into $M_{bol}$ we used
the relations by Bergeat, Knapik, \& Rutily (2002). Their Fig.~1
shows that the bolometric correction is well--defined for C-stars
with $J-K_s\lsim 2.1$, while for increasing colours the
uncertainty becomes larger. In our sample of 1079 stars, only 30
have $(J-K_s)_0 > 2.1$ and $flag(2)\leq 2$, these sources are
emphasized in Fig.~\ref{fig:logPMbol} as green triangles. These
stars have large amplitudes (see Fig.~\ref{fig:BJKcol}) which is
likely to influence their location in Fig.~\ref{fig:logPMbol}.

A comparison of the PL diagram with the models in Vassiliadis \&
Wood indicates that the bulk of C-rich LPVs in the SMC have lower
masses than 2.5-3~M$_\odot$, which at the metallicity of the SMC
corresponds to ages of 0.5-0.3~Gyr \citep{Pietrinferni+04}. This
agrees with the SFH derived by Harris \& Zaritsky (2004), who
found an active star formation during the past 3~Gyr with three
enhanced episodes at 2.5, 0.4, and 0.06~Gyr. Few carbon stars show
higher masses up to 5~M$_\odot$ corresponding to an age of
$\sim$0.1~Gyr.

\subsection{Spatial distribution}

The spatial distribution of stars in each sequence is
investigated. There seems to be no indication of a spatial
correlation between stars that occupy one or the other sequence in
the period--magnitude diagram with their location in the galaxy.
We also checked if there is any difference in the spatial
distribution of stars where different masses were expected. Again,
we do not find clear indication of a different distribution of
sources with a different initial mass. There are overall less
sources with high mass (or young), and a lack of these sources in
the Northeast, Northwest and Southwest regions compared to other
locations in the galaxy.  The distribution of sources with
intermediate and low mass is fairly similar.

\section{Summary and conclusions}
\label{section:conclusion}

This work analyzes and discusses the MACHO light-curves of 1079
C--stars. The sample consists of 751 photometrically selected
stars from the MC$^2$ catalogue according to $J-K_s\geq 1.33$~mag
and $K \leq 12$~mag criteria, and 328 spectroscopically confirmed
C-stars from RAW93 catalogue. Many of the photometrically selected
sample also have C-type spectra (RAW93), meaning that for 18\% of
all the sample we do not know their spectral type. However, given
the low metallicty of the SMC and a very low number of
spectroscopically known O-rich AGB stars in the SMC (G04), we
expect the contamination by O-rich AGB stars to be negligible. For
all the stars, we performed Fourier analysis of their
light--curves identifying up to 2 significant periodicities.

All the stars for which good quality light-curves exist were found
to vary with amplitudes of at least 0.05 mag in the two MACHO
photometric bands. The analysis was carried independently for
$B_M$ and $R_M$ light--curves, and the derived periods are
identical (to within the errors). After a selection based on the
quality of the light--curve, significance of the derived periods,
and quality of the periodicity fits, a total of 919 C--stars were
used in further analysis.

Carbon stars occupy bright parts of sequences $B$, $C$, and $D$ in
the $K_s - \log\,P$ diagram. None of the stars in our sample have
shorter periods characteristic of sequence $A$, which is in
agreement with recent studies by Fraser et al.\ (2005) in the LMC,
and Ita et al. (2004a,b), and G04 who observed this in the SMC as
well. The large majority of the stars have their primary
(dominant) periodicity on sequence $C$ or $D$, while more than 2/3
of sequence $B$ is populated by secondary periods of those stars
whose primary period is on sequences $C$ or $D$. The stars whose
primary period is on sequence $B$ are preferentially fainter and
bluer. This is in agreement with the models that predict change of
pulsation mode for the LPVs from higher overtones towards a
fundamental mode as they evolve along the AGB \citep[e.g.\
][]{Lattanzio&Wood03}.

Most of the stars with $J-K_s>2$ belong to sequence $C$, and only
a handful of these reddest variables show a long second period in
sequence $D$. This is a clue that dust obscuration cannot be a
cause of the long secondary periods \citep[see ][ for a detailed
discussion]{wood+04}. The luminosity functions of the three
sequences span a similar range of magnitudes, but the faint
distribution tail of sequence $D$ is more populated. Again, we
cannot explain it with the dust obscuration, as there are very few
stars redder than $J-K_s>2$ on this sequence. The width of this
sequence in the period--magnitude diagram is larger than that of
other sequences, especially at brighter magnitudes, but the lack
of stars with a period longer than ($log\ P\sim 3.5$) may be due
to  incomplete time coverage.

Stars belonging to different pulsational sequences are
homogenously mixed over the SMC area. Their masses were derived
from a comparison with the theoretical tracks of Vassiliadis \&
Wood (1993), and the majority of them are indicative of the major
star formation episode that took place $\sim 0.3-0.5$~Gyr ago.
This is in excellent agreement with Harris \& Zaritsky (2004), who
found enhanced star formation episodes at 2.5, 0.4, and 0.06 Gyr
in the SMC. However, our sample indicates a considerably weaker
star formation event at younger ages ($\sim 0.1$~Gyr ago), which
is expected as these younger, and thus more massive,  stars are
preferentially oxygen-rich, hence not in our sample. There is no
clear difference in the spatial distribution of the stars with
different masses.

The very long--time baseline of MACHO observations has allowed us
to confidently extract long periods up to ($log\ P\sim 3.5$). We
found that about 10\% of the variables fall on sequence $B$, 30\%
on $C$, and 34\% on $D$. The latter percentage is higher than the
25\% derived by Wood et al.\ (1999) and than the 21\% derived by
G04 in the LMC. It is also important to note that in the LMC,
Fraser et al.\ (2005) find only a small fraction of probable
C-stars with periods along sequence $D$. Monitoring of these
variables over more than an 8--year period photometrically and
spectroscopically is desirable in order to discern their nature.
According to Wood et al.~(2004) these stars belong to the only
class of bright large amplitude variables whose properties cannot
be explained with theoretical models at present.

\appendix

\section{Determination of Periods and Amplitudes}
\subsection{Method}
\label{section:method}

\begin{figure*}[t]
\center
\includegraphics[width=14cm]{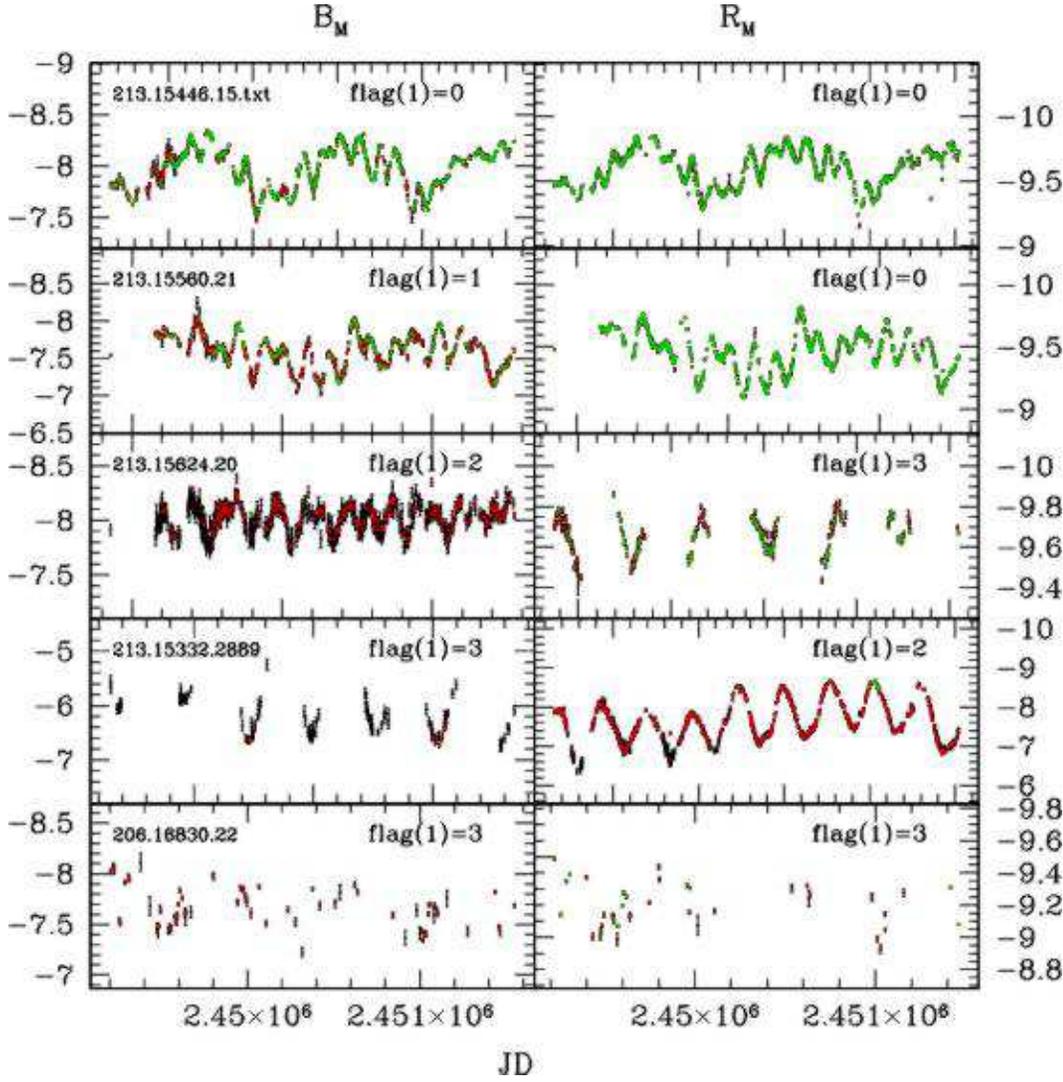}
   \caption{Examples of light--curves with
   different values of the data--quality $flag(1)$. Data with
   photometric errors $ 0.05 \leq \sigma_X < 0.1$ are plotted as black dots;
   data with $0.01 \leq \sigma_X < 0.05 $ are plotted as red squares; and data
   with $\sigma_X < 0.01 $ as green crosses. Left (right) panels
   illustrate $B_M$ ($R_M$) photometry. See the electronic edition of
   the Journal for a colour version of the figure.}
   \label{fig:lc_flag1}
\end{figure*}

\begin{figure*}[t]
\center{
\includegraphics[width=14cm]{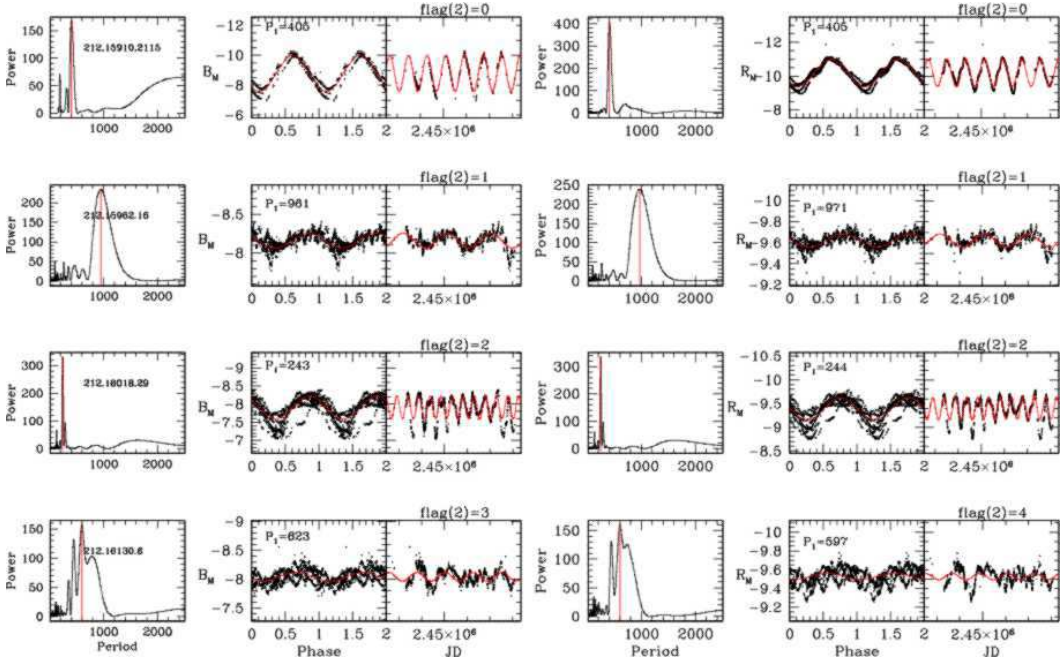}
   \caption{Examples of light--curves with
   different values of $flag(2)$. Left (right) panels illustrate $B_M$
   ($R_M$) photometry. For each band we show the periodogram and the
   light--curve in the phase and time domain. See the electronic
   edition of the Journal for a colour version of the figure.}
   \label{fig:lc_flag2}}
\end{figure*}

The method used to determined the first period is already
discussed in Rejkuba et al.\ (2003). To detect the second period
we used the  following fitting function:

\begin{eqnarray}
\label{secondperiod}
X(t)& = & A_1 \cos \left( 2 \pi \frac{(t - t_0)}{P_1} \right) +
B_1 \sin \left( 2 \pi \frac{(t - t_0)}{P_1} \right) \\
  & + &
A_2 \cos \left( 2 \pi \frac{(t - t_0)}{P_2} \right) +
B_2 \sin \left( 2 \pi \frac{(t - t_0)}{P_2} \right) \nonumber \\
& + & X_0 \nonumber
\end{eqnarray}
where the first guess for the second period comes from the second,
or sometimes the third, strongest peak in the power spectrum.
Usually, the  first period is  obvious, while secondary peaks  are
less clear.  We noted  that  semiregular and  irregular  light
curves  show several peaks in their Fourier spectrum with similar
strength. Usually we followed the  strength sequences of the power
values to define the second period. However, the amplitudes  of
the peak are just a starting point. In  fact we  always inspected
the relative  results by eye, in order to check the reliability of
the second assigned period. In some cases we considered  the third
peak more appropriate  than the second, which could  be instead an
alias  (a false period  which seemingly fits the  data, as  well
as the correct  period).   The alias  frequencies $\nu_a$ are
related to the true frequency $\nu_t$ by $\nu_a =\nu_t \pm
n/(\Delta t)$, where $\Delta t$ is  the spacing in days between
the measurements. The pattern of the  aliases will therefore
depend on the true period and  the particular $\Delta t$ for the
star in question.

We call the  {\it first period} ($P_1$) the best fitting period
obtained  from the first term of Eq.~\ref{secondperiod} and ${\cal
P}_{1}$ the associated value of the power. We  call the  {\it
second  period} ($P_2$)  the  best fitting  period obtained from
Eq.~\ref{secondperiod} and ${\cal P}_{2}$ the associated value of
the power. Amplitudes ($A_{fit}$) associated with the first period
are defined from the sinusoidal fit as:

\begin{equation}
A_{fit}=2\times \sqrt{A_1^2+B_1^2}, \label{amplitude}
\end{equation}

\begin{figure*}[t]
\center{
\includegraphics[width=14cm]{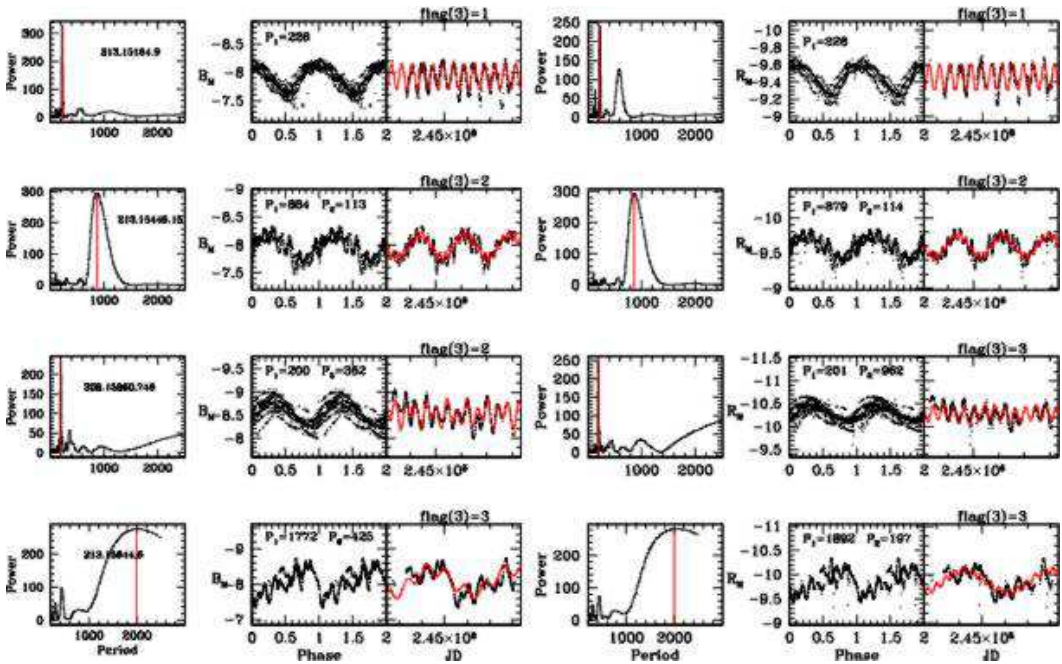}
   \caption{As in Fig.~\ref{fig:lc_flag2}, but for $flag(3)$.}
   \label{fig:lc_flag3}}
\end{figure*}

\subsection{Quality assessment}
\label{section:qualitycheck}

Table~\ref{tab:table_flags} summarizes the values of the three
quality flags associated to each light-curve and the number of
light-curves with a given value.  In particular, $flag(1)$ is
related  to the accuracy of the photometric measurements. We
defined $n_{0.05}$  and  $n_{0.01}$  as   the  ratio  between  the
number  of observations with,  respectively, magnitude error
$\sigma_X<0.05$ and $<0.01$  and the  total number  of
measurements  in  each light--curve (i.e.\  with  magnitude errors
$\Delta  < 0.1$).   Thus,  values  to $flag(1)$ are assigned as
follows:

\begin{eqnarray}
flag(1) & = & 0 \: \: \: if \: \: \:
        n_{0.01}\geq 0.7 \: \: \:  \: \: \: (excellent) \nonumber \\
flag(1) & = & 1 \: \: \: if \: \: \: 0.5 \leq n_{0.05}< 1 \: \: \: and \: \: \: 0.1 \leq n_{0.01}<0.7 \: \: \:  \nonumber \\
        &   &   \: \: \: (good); \nonumber \\
flag(1) & = & 2 \: \: \: if \: \: \: 0.5 \leq n_{0.05}< 1 \: \: \: and \: \: \: n_{0.01}<0.1 \: \: \:  \nonumber \\
        &   &   \: \: \:  (fair); \nonumber \\
flag(1) & = & 3 \: \: \: if \: \: \: n_{0.05}< 0.5 \: \: \: and \: \: \: n_{0.01}<0.05 \: \: \: \nonumber \\
        &   &   \: \: \:  (few \: or \: noisy \: data) \nonumber
\end{eqnarray}

A  sample  of  light--curves  of  different types,  together  with
the associated   $flag(1)$,  is   shown  in
Fig.~\ref{fig:lc_flag1}.  The algorithm works  well; only for a
handful of cases  the inspection by eye has revealed that data are
sparse over several magnitudes even if they have very  small
magnitude error. In these  cases we re--assigned $flag(1)=3$. The
majority of light--curves  classified as $flag(1)=2$ look like
213.15624.20      $B_M$-light       curve      in
Fig.~\ref{fig:lc_flag1}. Only  in a few  cases are very regular
and large amplitude  LPVs classified with  $flag(1)=2$ or even 3
(see for example   213.15332.2889,  $B_M$,   and  $R_M$
light--curve   in  the Fig.~\ref{fig:lc_flag1}).

The fit quality flag  ($flag(2)$) and the periodicity flag
($flag(3)$; see  Table~\ref{tab:table_flags})  are less  objective
and much  more related to  an inspection  by eye than  $flag(1)$.
In order  to assign values  to these flags,  first we  carefully
eye--inspected  the first period in the phase and  time diagrams.
Then, we looked at the combined periodicities
(Eq.~\ref{secondperiod}).  On the basis of  these inspections we
assigned  values for $flag(2)$ and   $flag(3)$ according   to
Table~\ref{tab:table_flags}.    In Fig.~\ref{fig:lc_flag2} we plot
light--curves with different values of $flag(2)$    referring only
to their    first periodicity. Fig.~\ref{fig:lc_flag3} illustrates
examples of light--curves  with different values of $flag(3)$. In
both figures for  each star we plot the corresponding power
spectrum and the light curve in phase and time domain and overplot
the best fitting sinusoidal function.

\begin{figure}[h]
\center
   \includegraphics[width=9cm]{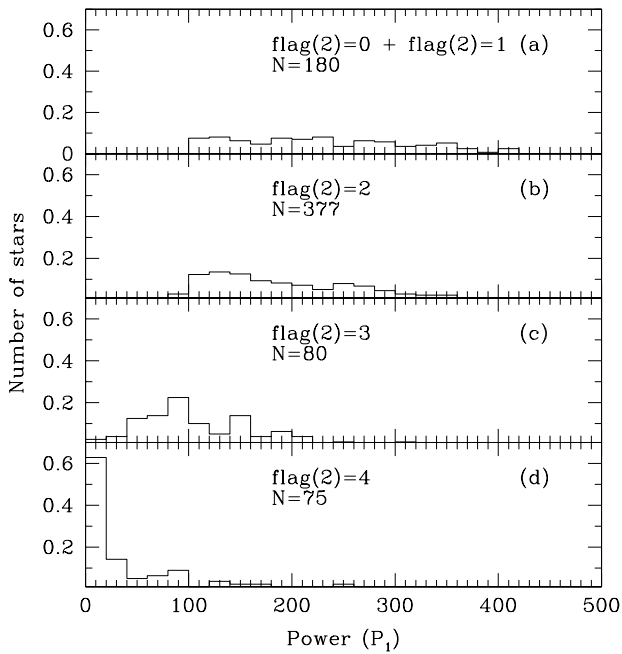}
   \caption{Histogram of the power of the first period ($B_M$--light
   curves) as a function of the defined fit-quality flag. The relative
   number of stars is indicated in each panel.}  \label{fig:hist_flag2}
\end{figure}

\begin{figure}[h]
\center
\includegraphics[width=9cm]{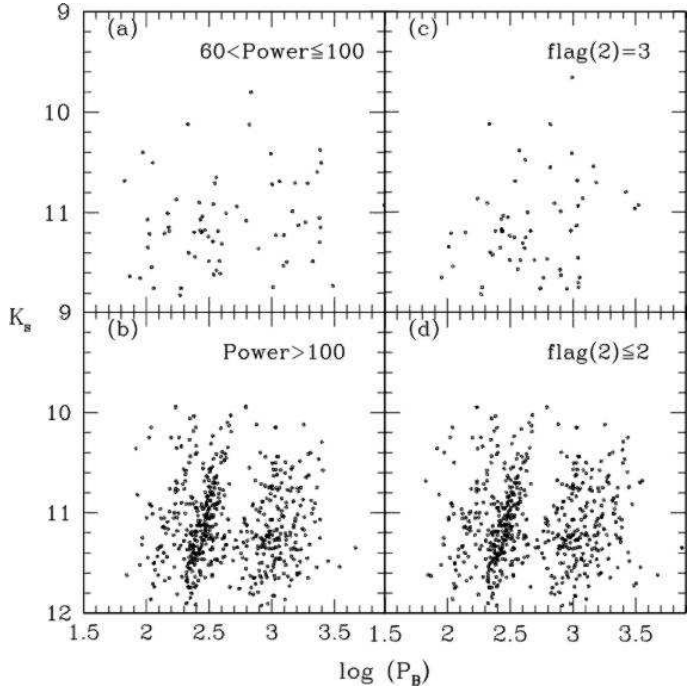}
   \caption{Distributions of the first period ($B_M$--light--curves) as
   a function of the power value (left panels) and $flag(2)$ (right
   panels).}  \label{fig:fig_flag2}
\end{figure}

In order to provide an  automatic classification of the fit
quality, we analyzed  the correlation between  $flag(2)$ and  the
strength  of the power   of  the   first  period   (${\cal  P}_1$)
as   follows.   In Fig.~\ref{fig:hist_flag2}  the histogram of
${\cal P}_1$  is reported for  different values  of $flag(2)$.  It
is  clear that  starting from $flag(2)=0$ to  $flag(2)=4$ the peak
of the  distribution moves toward lower   power  values.  In
particular,  for   $flag(2)=  0$   and  1
(Fig.~\ref{fig:hist_flag2}$a$) the only  region populated is that
with ${\cal P}_1  \geq 100$;  the bulk of  variables with
$flag(2)=2$ have ${\cal P}_1 \gsim  100$ with only a small number
of stars with ${\cal P}_1 < 100$.  The distribution of variables
with  $flag(2)=3$ peaks in the  range  $80  \leq {\cal  P}_1  <
100$,  and  those for  which  no satisfactory period could be
derived mainly have ${\cal P}_1 < 40$.

\begin{figure}[h]
\center
\includegraphics[width=9cm]{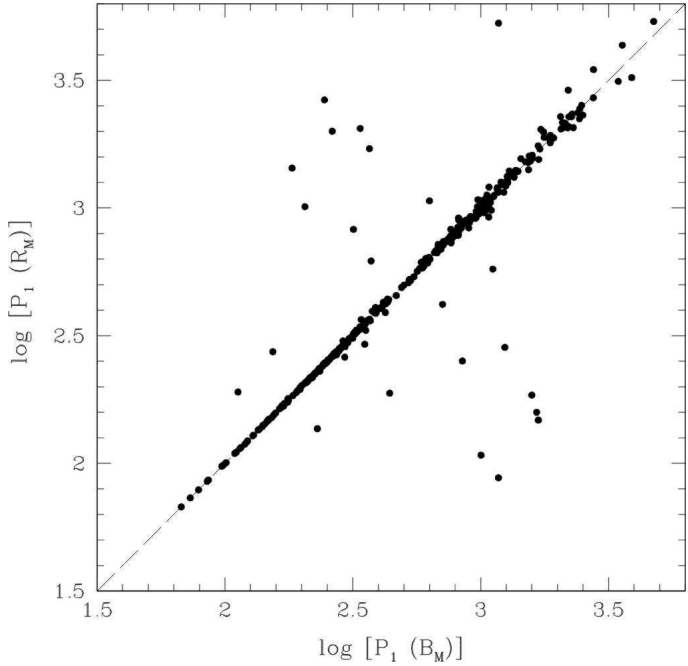}
   \caption{Relation between the first periods derived from $B_M$ and $R_M$
   light--curves.
    }
   \label{fig:PBR}
\end{figure}

Figure \ref{fig:fig_flag2} emphasizes  the good correlation
between these two quantities. In panels ($a,b$) we plot $log(P)$
vs. $K_s$ (see next section for a discussion) for stars with $60 <
{\cal P}_1 < 100$ ($a$) and stars with ${\cal P}_1 > 100$ ($b$).
Panels ($c,d$) show the same quantities but for stars with
$2flag(2)=3$ $(c)$, and $flag(2)  \leq 2$  ($d$).  We conclude
that  the selection described in  panels ($b,d$) outlines three
sequences and the distribution of stars in both panels are very
similar. These results
 are  related to $B_M$  light--curves, however an  inspection of
$R_M$  light--curves gives  similar  results. Fig.~\ref{fig:PBR}
shows an  almost perfect correlation  between the first period
obtained  from $B_M$ and $R_M$  light--curves.  Only  stars with
$flag(2)\leq 2$  are plotted. For 94\% of the objects the residual
dispersion of the differences is lower than 4\%. Only a few points
scatter away from  the 1:1 line, implying that the primary period
is the same in both  bands. Most of these points occur because the
first period (here ${\cal P}_1$) found from the $B_M$ light--curve
corresponds to the second period (${\cal P}_2$) found from the
$R_M$  light--curve, and confronting only ${\cal P}_1$ between
both light--curves produces a point that does not follow the
correlation. A few other points occur when the period ${\cal P}_1$
derived from the $B_M$ light--curve is twice that derived from the
 $R_M$  light--curve.

Summarizing, an automatic selection criterion (i.e. by using
${\cal P}_1$) can  be applied with a good level  of confidence for
the analysis of  the primary periods of a large sample  of
variables.

To define $flag(3)$ in a similar automatic way as $flag(2)$, we
considered the values of the reduced $\chi^2$ of  the sinusoidal
fit.  In most  cases the $\chi ^2$ value decreases  when two
periodicities are considered,  even in cases we classified with
$flag(3)=1$, i.e.\ when only one period could be reliably derived.
Perhaps this indicates the complexity of stellar pulsation.  For
these stars $flag(3)=1$ does  not mean that the  light--curve is
clearly characterized  by one  periodicity only; rather  it
implies that the second period is much less regular and could not
be fitted well with a set of sinusoid functions.

\subsection{Comparison between photometric amplitude and amplitude of
the sinusoidal fit}

\begin{figure}
\center
  \includegraphics[width=9cm]{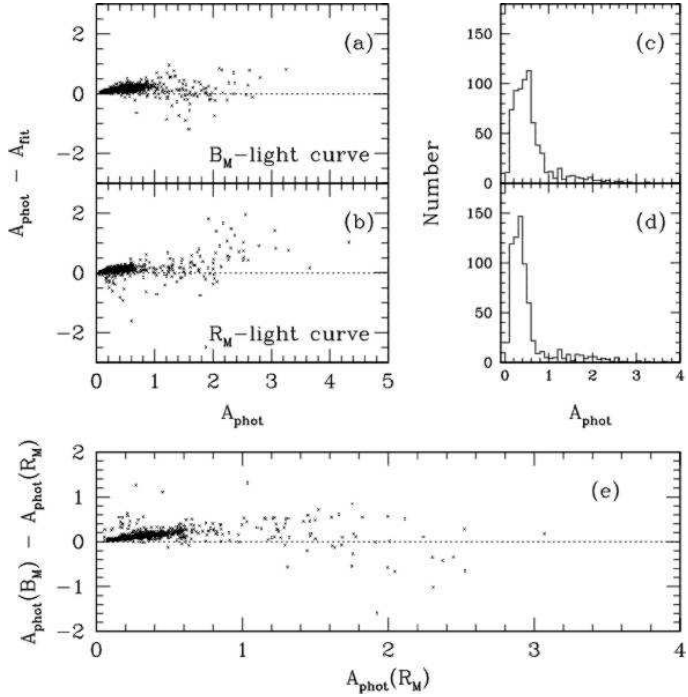}
   \caption{{\bf Panel $(a,b)$}:
   The residual $(A_{phot}-A_{fit})$ is plotted against
   $A_{phot}$. {\bf Panel $(c,d)$}: Histogram of the amplitude is reported for
   $B_M$ ($c$) and $R_M$ ($d$) magnitudes.
   {\bf Panel $(e)$}: Difference between amplitude from $R_M$- and $B_M$-light
    curves.   Only stars with $flag(2)<3$ are shown.}
   \label{fig:amplitude}
\end{figure}

Figure  \ref{fig:amplitude}  shows the  relation  between the
pulsation amplitude derived  from the sinusoidal  fit ($A_{fit}$)
and  that from the peak--to--peak magnitude difference
($A_{phot}$)  for $\bar {B}_M$ and $\bar {R}_M$ light--curves
separately.  The  latter  value  is defined  as $A_{phot}=\bar
{B}_M(max)-\bar {B}_M(min)$,  where  $\bar {B}_M(max)$ and $\bar
{B}_M(min)$ are the  maximum and the minimum values averaged over
a few data  points in  order to  avoid spurious  detections. For
sources  of  regular  periodicity  with excellent  observational
data (i.e. low  photometric errors and $flag(1)=0$) that  clearly
show only one periodicity,  we discarded the  first 5 measurements
and  took the mean over the brighter (fainter) ten measurements
(see for example the light--curve  212.15910.2115  in
Fig.~\ref{fig:lc_flag2}). Since  the majority  of sources show
less regular  light--curves with  bumps and multiperiodicity, the
first  brighter (fainter) 150 measurements, which could be
disturbed by the  secondary periodicity with $P\sim 150$ days, are
excluded. Then,  we computed the  ``maximum'' (``minimum'')  value
as the average  over the  next brighter (fainter) 40 photometric
measurements.

The  sinusoidal fit  clearly predicts  smaller amplitudes  compared to
$A_{phot}$.  Quantitatively, $A_{phot}(B_M)-A_{fit}(B_M)
\simeq 0.12 \pm 0.22$  and $A_{phot}(R_M)-A_{fit}(R_M) \simeq 0.08 \pm
0.25$. This  is expected  in case of  deviations from  pure sinusoidal
variability, due  to the presence  of irregularities in  the intrinsic
light  variation and photometric  uncertainties caused  by measurement
errors and variations  in seeing. The latter may  lead to blending and
spurious measurements due  to the presence of cosmic  rays and defects
on the CCD.  As can be seen from  Fig.~\ref{fig:lc_flag2} this applies
to all but the most regular variables with $flag(2)=0$.

The histograms of  amplitudes in  the two  MACHO  photometric
bands are reported  in Fig.~\ref{fig:amplitude}$c,d$.   The bulk
of sources  have  $0.2 \leq A_{phot}(B_M)  \leq  0.5  $  and  $0.1
\leq  A_{phot}(B_M)  \leq  0.4 $. As expected, amplitudes  in
$B_M$  are slightly higher than those in  $R_M$ (see
Fig.~\ref{fig:amplitude}$e$). The  difference  in  the  mean
amplitude of  the two bands  is $A_{phot}(B_M)-A_{phot}(R_M) =
0.15\pm0.18$.

\begin{acknowledgements}
    We thank the anonymous referee for providing
    constructive and insightful comments that have greatly
    improved the paper.
    We thank N. Delmotte for giving
    us the $\mathrm{MC}^2$ catalogue of SMC data before publication.
    Support for this project was provided by the ESO Director
      General Discretionary Fund.
    The support given by ASTROVIRTEL, a Project funded by the European
    Commission under FP5 Contract No. HPRI-CT-1999-00081, is
    acknowledged.
    This paper utilizes public domain data originally obtained by
    the MACHO Project, whose work was performed under the joint
    auspices of the U.S. Department of Energy, National Nuclear
    Security Administration by the University of California,
    Lawrence Livermore National Laboratory under contract
    No. W-7405-Eng-48, the National Science Foundation through the
    Center for Particle Astrophysics of the University of
    California under cooperative agreement AST-8809616, and the
    Mount Stromlo and Siding Spring Observatory, part of the
    Australian National University.  It also makes use of
    data products from the Two Micron All Sky Survey, which is a
    joint project of the University of Massachusetts and the
    Infrared Processing and Analysis center/California Institute
    of Technology, funded by the National Aeronautics and Space
    Administration and the National Science Foundation.
    The DENIS project is partially funded by European
    Commission through SCIENCE and Human Capital and Mobility plan
    grants. It is also supported, in France by the Institut
    National des Sciences de l'Univers, the Education Ministry and
    the Centre National de la Recherche Scientifique, in Germany
    by the State of Baden-W\"urtemberg, in Spain by the DG1CYT, in
    Italy by the Consiglio Nazionale delle Ricerche, in Austria by
    the Fonds zur F\"orderung der wissenschaftlichen Forschung und
    Bundesministerium fuer Wissenschaft und Forschung, in Brazil by
    the Foundation for the development of Scientific Research of
    the State of Sao Paulo (FAPESP), and in Hungary by an OTKA
    grant and an ESOC\&EE grant.
\end{acknowledgements}

\end{document}